\documentclass[a4paper,11pt]{article}
\pdfoutput=1 
\usepackage{jcappub}
\usepackage{amsmath}
\usepackage{amssymb}
\usepackage{bm}
\usepackage{graphicx}
\usepackage{enumitem}
\usepackage{geometry}
\usepackage{xspace}
\usepackage{pdflscape}
\usepackage{enumerate}

\makeatletter
\gdef\@fpheader{}
\g@addto@macro\bfseries{\boldmath}
\makeatother

\newcommand{\ie}{{i.e.~}}

\newcommand{\eg}{e.g.~}

\newcommand{\E}{E_{\mathrm{B}}}

\let\oldsqrt\sqrt
\def\sqrt{\mathpalette\DHLhksqrt}
\def\DHLhksqrt#1#2{%
\setbox0=\hbox{$#1\oldsqrt{#2\,}$}\dimen0=\ht0
\advance\dimen0-0.2\ht0
\setbox2=\hbox{\vrule height\ht0 depth -\dimen0}%
{\box0\lower0.4pt\box2}}

\newcommand{\order}[1]{\mathcal{O}\!\left(#1\right)}

\newcommand{\dd}{\mathrm{d}}
\newcommand{\ee}{e}

\newcommand{\sss}[1]{{\scriptscriptstyle{#1}}}

\newcommand{\uPl}{\mathrm{Pl}}
\newcommand{\uin}{\mathrm{in}}

\newcommand{\usssPl}{\sss{\uPl}}

\newcommand{\calH}{\mathcal{H}}

\newcommand{\Mp}{M_\usssPl}

\newcommand{\efolds}{$e$-folds}

\newcommand{\beq}{\begin{equation}}
\newcommand{\eeq}{\end{equation}}
\newcommand{\bea}{\begin{equation}\begin{aligned}}
\newcommand{\eea}{\end{aligned}\end{equation}}

\newlength{\wsingfig}
\newlength{\wdblefig}
\newlength{\wquadfig}
\newlength{\wtriplefig}
\newcommand{\Eq}[1]{Eq.~(\ref{#1})}
\newcommand{\Eqs}[1]{Eqs.~(\ref{#1})}
\newcommand{\Fig}[1]{Fig.~{\ref{#1}}}

\newcommand{\Ref}[1]{Ref.~{\cite{#1}}}

\newcommand{\Sec}[1]{Sec.~\ref{#1}}
\newcommand{\Secs}[1]{Secs.~\ref{#1}}
\newcommand{\App}[1]{Appendix~\ref{#1}}

\subheader{}

\title{Stochastic inflation beyond slow roll}

\author[a]{Chris Pattison,}
\author[b,a]{Vincent Vennin,}
\author[a,c]{Hooshyar Assadullahi,}
\author[a]{and David Wands}

\affiliation[a]{Institute of Cosmology \& Gravitation, University of Portsmouth, Dennis Sciama Building, Burnaby Road, Portsmouth, PO1 3FX, United Kingdom}
\affiliation[b]{Laboratoire Astroparticule et Cosmologie, Universit\'e Denis Diderot Paris 7, 75013 Paris,
France}
\affiliation[c]{School of Mathematics and Physics, University of Portsmouth, Lion Gate Building, Lion Terrace, Portsmouth, PO1 3HF, United Kingdom}

\emailAdd{christopher.pattison@port.ac.uk}
\emailAdd{vincent.vennin@port.ac.uk}
\emailAdd{hooshyar.assadullahi@port.ac.uk}
\emailAdd{david.wands@port.ac.uk}

\date{today}

\begin{document}

\sloppy

\abstract{We study the stochastic formalism of inflation beyond the usual slow-roll approximation. We verify that the assumptions on which the stochastic formalism relies still hold even far from the slow-roll attractor. This includes demonstrating the validity of the separate universe approach to evolving long-wavelength scalar field perturbations beyond slow roll. We also explain that, in general, there is a gauge correction to the amplitude of the stochastic noise. This is because the amplitude is usually calculated in the spatially-flat gauge, while the number of e-folds is used as the time variable (hence one works in the uniform-$N$ gauge) in the Langevin equations. We show that these corrections vanish in the slow-roll limit, but we also explain how to calculate them in general. We compute them in difference cases, including ultra-slow roll and the Starobinsky model that interpolates between slow roll and ultra-slow roll, and find the corrections to be negligible in practice. This confirms the validity of the stochastic formalism for studying quantum backreaction effects in the very early universe beyond slow roll.}



\maketitle

\section{Introduction}
\label{sec:intro}

The leading paradigm for the very early universe is that of cosmological inflation \cite{Starobinsky:1980te, Sato:1980yn, Guth:1980zm, Linde:1981mu, Albrecht:1982wi, Linde:1983gd}, which describes a phase of primordial high-energy accelerated expansion. While inflation was proposed as a solution to the classical problems of the standard hot big bang cosmology, its most impressive achievement is that it provides the seeds for large-scale structure, through quantum vacuum fluctuations that become amplified during inflation \cite{Mukhanov:1981xt, Mukhanov:1982nu, Starobinsky:1982ee, Guth:1982ec, Hawking:1982cz, Bardeen:1983qw}.

As these fluctuations are stretched to cosmological distances, they backreact on the local background expansion rate. This quantum backreaction effect can be modelled through a stochastic formalism known as stochastic inflation~\cite{Starobinsky:1982ee, Starobinsky:1986fx}. This approach is usually studied in the slow-roll regime, which is an attractor of the classical field dynamics, where the frictional force due to the expansion balances the force coming from the potential gradient. In this regime, excellent agreement between stochastic inflation and usual quantum field theoretic techniques has been found~\cite{Starobinsky:1994bd, Tsamis:2005hd, Finelli:2008zg, Garbrecht:2013coa, Vennin:2015hra, Onemli:2015pma, Burgess:2015ajz, Vennin:2016wnk, Hardwick:2017fjo, Tokuda:2017fdh}.

Recently, situations in which non-slow-roll stochastic effects are at play have been highlighted~\cite{Garcia-Bellido:2017mdw, Germani:2017bcs, Firouzjahi:2018vet, Biagetti:2018pjj, Ezquiaga:2018gbw}. For instance, if the inflationary potential features a very flat section close to the end of inflation, large curvature perturbations could be produced that later collapse into primordial black holes. If such a flat portion exists, it may be associated with both large stochastic diffusion~\cite{Pattison:2017mbe} and deviations from slow-roll, \eg along the so-called ultra-slow-roll (or ``friction dominated'') regime~\cite{Inoue:2001zt, Kinney:2005vj}, which in some cases can be stable~\cite{Pattison:2018bct}. This explains the need for implementing the stochastic inflation programme beyond slow roll, which is the aim of the present work.

This paper is organised as follows. In \Sec{sec:stochastic}, we quickly review the stochastic inflation formalism and identify the three main requirements for the validity of this approach: the quantum-to-classical transition of super-Hubble fluctuations, the validity of the separate universe approach, and the consistent implementation of gauge corrections. The two latter requirements are the non-trivial ones and we examine them in \Secs{sec:separateuniverse} and~\ref{sec:uniformexpansion} respectively. Although recently questioned~\cite{Cruces:2018cvq}, we find the separate universe approach to hold beyond slow roll, and we explain how the gauge corrections to the amplitude of the stochastic noise (that vanish in the slow-roll regime if the number of \efolds~is used as a time variable) can be derived in general. We then apply this program to three situations of interest: slow roll in \Sec{sec:slowroll}, where we recover the usual results, ultra-slow roll in \Sec{sec:USR}, and the Starobinsky model in \Sec{sec:starobinsky}, which interpolates between an ultra-slow-roll and a slow-roll phase. In all cases, we find the gauge corrections to be negligible, allowing for the usual stochastic formulation to be employed. 
\section{Requirements for stochastic inflation}
\label{sec:stochastic}
The action describing 4-dimensional gravity in a curved space-time, with metric $g_{\mu\nu}$, minimally coupled to a scalar field $\phi$, reads
\bea \label{eq:4d:action}
S = \int \dd^4x\sqrt{-g}\left[ \frac{\Mp^2}{2}R - \frac{1}{2}g^{\mu\nu}\partial_{\mu}\phi\partial_{\nu}\phi - V(\phi) \right] \, .
\eea 
In this work, we consider a single inflaton field $\phi$ with potential $V(\phi)$ for simplicity, but our results can easily be extended to multiple-field setups. We also restrict our analysis to scalar fluctuations only, and expand the metric about the flat Friedmann-Lema\^itre-Robertson-Walker (FLRW) line element,
\bea \label{eq:perturbedlineelement:FLRW}
\dd s^2 &= -(1+2A)\dd t^2 + 2a\partial_{i}B\dd x^{i}\dd t + a^2\left[ (1-2\psi)\delta_{ij} + 2\partial_i \partial_j E \right]\dd x^i \dd x^j \, ,
\eea 
where $a$ is the scale factor of the universe. 

Away from slow roll, the homogeneous background field $\phi$ and its conjugate momentum $\pi$ are two independent dynamical variables and stochastic inflation needs to be formulated in the full phase space (see \Ref{Grain:2017dqa} for a more detailed analysis). This can be done by deriving the Hamiltonian equations from the action \eqref{eq:4d:action},
\begin{align} 
\label{eq:conjmomentum} \frac{\partial \hat{\phi}}{\partial N} &= \hat{\pi} \, ,\\
\label{eq:KG:efolds} \frac{\partial\hat{\pi}}{\partial N} &+ \left(3-\epsilon_{1}\right)\hat{\pi} + \frac{V_{,\phi}(\hat{\phi})}{H^2} = 0 \, ,
\end{align}
where $N=\ln a$ is the number of \efolds, $\epsilon_{1}\equiv - \dot{H}/H^2$ is the first slow-roll parameter, $H\equiv \dot{a}/a$ is the Hubble parameter, a dot denotes derivatives with respect to cosmic time $t$ and a subscript $_{,\phi}$ means derivative with respect to the field $\phi$. At this stage, $\hat{\phi}$ and $\hat{\pi}$ are quantum operators, as stressed by the hats. The Hubble parameter is related to the field phase-space variables through the Friedmann equation,
\bea 
\label{eq:friedmann}
H^2 = \dfrac{V+\dfrac{\dot{\phi}^2}{2}}{3\Mp^2} \, ,
\eea 
where $\Mp$ is the reduced Planck mass.

When linear fluctuations are added to the homogenous field and its conjugate momentum, they can be split according to 
\begin{align}
\hat{\phi} &= \hat{\bar{\phi}} + \hat{\phi}_{\mathrm{s}} \label{eq:decomp:phi}\, , \\
\hat{\pi} &= \hat{\bar{\pi}} + \hat{\pi}_{\mathrm{s}} \label{eq:decomp:pi}\, ,
\end{align}
where $\hat{\phi}_{\mathrm{s}}$ and $\hat{\pi}_{\mathrm{s}}$ are the short-wavelength parts of the fields that can be written as 
\begin{align}
    \hat{\phi}_{\mathrm{s}} &= \int_{\mathbb{R}^3}\frac{\dd \bm{k}}{\left(2\pi\right)^{\frac{3}{2}}}W\left( \frac{k}{\sigma aH}\right) \left[ \ee^{-i\bm{k}\cdot\bm{x}}\phi_{\bm{k}}(N)\hat{a}_{\bm{k}} + \ee^{i\bm{k}\cdot\bm{x}}\phi_{\bm{k}}^{*}(N)\hat{a}^{\dagger}_{\bm{k}}  \right] \, ,\\
        \hat{\pi}_{\mathrm{s}} &= \int_{\mathbb{R}^3}\frac{\dd \mathbf{k}}{\left(2\pi\right)^{\frac{3}{2}}}W\left( \frac{k}{\sigma aH}\right) \left[ \ee^{-i\bm{k}\cdot\bm{x}}\pi_{\bm{k}}(N)\hat{a}_{\bm{k}} + \ee^{i\bm{k}\cdot\bm{x}}\pi_{\bm{k}}^{*}(N)\hat{a}^{\dagger}_{\bm{k}}  \right] \, .
\end{align}
In these expressions, $\hat{a}^{\dagger}_{\bm{k}}$ and $\hat{a}_{\bm{k}}$ are creation and annihilation operators, and $W$ is a window function that selects out modes such that  $k/(\sigma aH) > 1$, where $\sigma\ll 1$ is the coarse-graining parameter. The coarse-grained fields $\bar{\phi}$ and $\bar{\pi}$ thus contain all wavelengths that are much larger than the Hubble radius, $k< \sigma aH$. They stand for the local background values of the fields, that are continuously perturbed by the small wavelength modes, as they emerge from $ \hat{\phi}_{\mathrm{s}}$ and $ \hat{\pi}_{\mathrm{s}}$ and cross the coarse-graining radius. 

Inserting the decompositions \eqref{eq:decomp:phi} and \eqref{eq:decomp:pi} into the classical equations of motion \eqref{eq:conjmomentum} and \eqref{eq:KG:efolds}, to linear order in the short-wavelength parts of the fields, the equations for the long-wavelength parts become
\begin{align}
\frac{\partial \hat{\bar{\phi}}}{\partial N} &= \hat{\bar{\pi}} + \hat{\xi}_{\phi}(N) \label{eq:conjmomentum:langevin:quantum} \, ,\\
\frac{\partial \hat{\bar{\pi}}}{\partial N} &= -\left(3-\epsilon_{1}\right)\hat{\bar{\pi}} - \frac{V_{,\phi}(\hat{\bar{\phi}})}{H^2} +\hat{\xi}_{\pi}(N) \label{eq:KG:efolds:langevin:quantum} \, ,
\end{align}
where the source functions $\hat{\xi}_{\phi}$ and $\hat{\xi}_{\pi}$ are given by
\begin{align}
    \hat{\xi}_{\phi} &= -\int_{\mathbb{R}^3}\frac{\dd \mathbf{k}}{\left(2\pi\right)^{\frac{3}{2}}}\frac{\dd W}{\dd N}\left( \frac{k}{\sigma aH}\right) \left[ \ee^{-i\bm{k}\cdot\bm{x}}\phi_{\bm{k}}(N)\hat{a}_{\bm{k}} + \ee^{i\bm{k}\cdot\bm{x}}\phi_{\bm{k}}^{*}(N)\hat{a}^{\dagger}_{\bm{k}}  \right]\, , \\
    \hat{\xi}_{\pi} &= -\int_{\mathbb{R}^3}\frac{\dd \mathbf{k}}{\left(2\pi\right)^{\frac{3}{2}}}\frac{\dd W}{\dd N}\left( \frac{k}{\sigma aH}\right) \left[ \ee^{-i\bm{k}\cdot\bm{x}}\pi_{\bm{k}}(N)\hat{a}_{\bm{k}} + \ee^{i\bm{k}\cdot\bm{x}}\pi_{\bm{k}}^{*}(N)\hat{a}^{\dagger}_{\bm{k}}  \right] \, .
\end{align}
If the window function is taken to be a Heaviside function, then the two-point correlation functions of the sources are given by
\bea 
\langle 0| \hat{\xi}_{\phi}(N_1)\hat{\xi}_{\phi}(N_2)|0\rangle & 
\kern-0.1em = \kern-0.1em
 \frac{1}{6\pi^2}\frac{\dd k_\sigma^3(N)}{\dd N}\bigg|_{N_1}\left\vert\phi_{\bm{k}_\sigma}(N_1) \right\vert^2 \delta\left(N_1 - N_2\right) \kern-0.1em ,\\
\langle 0| \hat{\xi}_{\pi}(N_1) \hat{\xi}_{\pi}(N_2)|0\rangle &
\kern-0.1em = \kern-0.1em
\frac{1}{6\pi^2}\frac{\dd k_\sigma^3(N)}{\dd N}\bigg|_{N_1}\left\vert \pi_{\bm{k}_\sigma}(N_1) \right\vert^2 \delta\left(N_1 - N_2\right) \kern-0.1em, \\
\langle 0| \hat{\xi}_{\phi}(N_1)\hat{\xi}_{\pi}(N_2)|0\rangle &
\kern-0.1em = \kern-0.1em \langle 0| \hat{\xi}_{\pi}(N_1) \hat{\xi}_{\phi}(N_2)|0\rangle^* 
\kern-0.1em = \kern-0.1em
 \frac{1}{6\pi^2}\frac{\dd k_\sigma^3(N)}{\dd N}\bigg|_{N_1}\phi_{\bm{k}_\sigma}(N_1) \pi^*_{\bm{k}_\sigma}(N_1)  \delta\left(N_1 \kern-0.1em - \kern-0.1em N_2\right)  \kern-0.1em,
 \label{eq:noise:correlators}
\eea 
where $k_\sigma \equiv \sigma a H$ is the comoving coarse-graining scale. The idea is then to view the source functions as random Gaussian noises rather than quantum operators, correlated according to \Eqs{eq:noise:correlators}, and to interpret \Eqs{eq:conjmomentum:langevin:quantum} and~(\ref{eq:KG:efolds:langevin:quantum}) as stochastic Langevin equations for the random field variables $\bar{\phi}$ and $\bar{\pi}$, 
\begin{align}
\frac{\partial {\bar{\phi}}}{\partial N} &= {\bar{\pi}} + {\xi}_{\phi}(N) \label{eq:conjmomentum:langevin} \, ,\\
\frac{\partial {\bar{\pi}}}{\partial N} &= -\left(3-\epsilon_{1}\right){\bar{\pi}} - \frac{V_{,\phi}({\bar{\phi}})}{H^2(\bar{\phi},\bar{\pi})} +{\xi}_{\pi}(N) \label{eq:KG:efolds:langevin} \, ,
\end{align}
where we have removed the hats to stress that we now work with stochastic quantities rather than quantum operators. The validity of such an approach relies on three main requirements:
\begin{itemize}
\item {\em quantum-to-classical transition}

The replacement of quantum operators by stochastic fields is a non-trivial procedure. For instance, stochastic variables always commute while quantum operators do not. From the last of \Eqs{eq:noise:correlators}, one can see that this notably implies the imaginary part of $\phi_{\bm{k}}\pi^*_{\bm{k}}$ to be negligible compared to its real part. During inflation, cosmological perturbations are placed in a two-mode highly squeezed state on large scales and indeed experience such a quantum-to-classical transition~\cite{Polarski:1995jg, Lesgourgues:1996jc, Kiefer:2008ku}. The only requirement for this to happen is the dominance of a growing mode over a decaying mode, which is guaranteed as long as perturbations get amplified outside the Hubble radius. This does not rely on slow roll and is therefore not a problem in general.

Furthermore, the importance of this first requirement should be taken with a grain of salt. First, hermitian two-point functions involving the field and its conjugate momentum, or any higher-order correlator involving only one phase-space variable, can be well reproduced by a stochastic description regardless of the amount of quantum squeezing~\cite{Martin:2015qta, Grain:2017dqa}. Second, the amount of squeezing is defined up to a choice of phase-space canonical variables, and can be set to any arbitrary value by performing a suitable canonical transformation~\cite{Grain:2017dqa}. Moreover, the real part of $\phi_{\bm{k}}\pi^*_{\bm{k}}$ can be set to zero after a suitable rotation in phase space, so the above classical criterion is not invariant under canonical transformations. Third, there are a class of observables called ``improper''~\cite{2005PhRvA..71b2103R}, the expectation values of which can never be reproduced by a stochastic theory, even in the large-squeezing limit (giving rise \eg to Bell inequality violations~\cite{Martin:2016tbd, Martin:2017zxs}). 

The quantum-to-classical transition is therefore a delicate concept, which however does not rely on the slow-roll approximation, hence does not hinder the use of a stochastic formalism beyond this regime. 

\item {\em separate universe approach}

Since the spatial gradients in the Langevin equations~(\ref{eq:conjmomentum:langevin}) and~(\ref{eq:KG:efolds:langevin}) are neglected, one assumes that, on super-Hubble scales, each Hubble patch evolves forward in time independently of the other patches, and under a locally FLRW metric. This is the so-called separate universe picture~\cite{Salopek:1990jq, Sasaki:1995aw, Wands:2000dp, Lyth:2003im, Rigopoulos:2003ak, Lyth:2005fi}, or quasi-isotropic \cite{Lifshitz:1960, Starobinsky:1982mr, Comer:1994np, Khalatnikov:2002kn} picture. The validity of this approximation beyond slow roll has recently been questioned in \Ref{Cruces:2018cvq}, and in \Sec{sec:separateuniverse}, we will show why it is in fact still valid.

\item {\em use of the uniform-$N$ gauge}

In order to derive the Langevin equations~(\ref{eq:conjmomentum:langevin}) and~(\ref{eq:KG:efolds:langevin}), only the field variables have been perturbed according to \Eqs{eq:decomp:phi} and~(\ref{eq:decomp:pi}), and not the entries of the metric. In particular, the lapse function, \ie $A$ in the notation of \Eq{eq:perturbedlineelement:FLRW}, has been neglected. This implies that the Langevin equations are written in a specific gauge, namely the one where the time coordinate is fixed. Since we work with the number of \efolds~as the time variable, this corresponds to the uniform-$N$ gauge. In \Eqs{eq:noise:correlators}, the field perturbations $\phi_{\bm{k}}$ and $\pi_{\bm{k}}$ must therefore be calculated in that same gauge. However, it is common to compute the field perturbations in the spatially-flat gauge, since in that gauge, they are directly related to the gauge-invariant curvature perturbation, which is quantised in the Bunch-Davies vacuum. One must therefore compute the correction to the noise amplitude that comes from translating the field fluctuations in the spatially-flat gauge to the uniform-$N$ gauge, and this is what is done in detail in \Sec{sec:uniformexpansion}. 

Let us note that one could work with a different time coordinate, hence in a different gauge (for instance, working with cosmic time $t$ would imply working in the synchronous gauge). This is not a problem as long as one computes gauge-invariant quantities in the end, such as the curvature perturbation $\zeta$. However, since $\zeta$ is related to the fluctuation in the number of \efolds~in the so-called ``stochastic $\delta N$ formalism''~\cite{Fujita:2013cna, Vennin:2015hra}, we find it convenient to work with the number of \efolds~as a time variable. Another reason is that, as will be shown in \Sec{sec:uniformexpansion}, in the slow-roll regime, the spatially-flat gauge coincides with the uniform-$N$ gauge (but not, say, with the synchronous gauge), which makes the gauge correction identically vanish, and which explains why it is usually recommended~\cite{Vennin:2015hra} (but not compulsory) to work with $N$ as a time variable. 
\end{itemize}
\section{Separate universes}
\label{sec:separateuniverse}

The separate universe approach is valid when each causally-disconnected patch of the universe evolves independently, obeying the same field equations locally as in a homogeneous and isotropic (FLRW) cosmology. 
Combining \Eqs{eq:conjmomentum} and~(\ref{eq:KG:efolds}), the  Klein--Gordon equation for a homogeneous field in an FLRW cosmology, $\phi(t)$, is given by
\bea 
\label{eq:kleingordon}
\ddot{\phi} + 3 H\dot{\phi} + V_{,\phi} &= 0 \, .
\eea 

In this section, we derive the equation of motion for linear fluctuations about a homogeneous scalar field from (i) cosmological perturbation theory, and (ii) perturbations of the background FLRW equations of motion, i.e., the separate universe approach. We show that the two equations of motion match at leading order in a spatial gradient expansion, with or without slow roll.
\subsection{Cosmological perturbation theory}

At linear order in perturbation theory, the perturbed Klein-Gordon equation in Fourier space, with \Eq{eq:perturbedlineelement:FLRW}, gives~\cite{Gordon:2000hv, Malik:2008im}
\bea \label{eq:pertubations:general}
\ddot{\delta\phi_{\bm{k}}} + 3H\dot{\delta\phi_{\bm{k}}} + \left(\frac{k^2}{a^2}+V_{,\phi\phi}\right)\delta\phi_{\bm{k}} = -2V_{,\phi}A_{\bm{k}} + \dot{\phi}\left[ \dot{A_{\bm{k}}} + 3\dot{\psi_{\bm{k}}} + \frac{k^2}{a^2}\left(a^2\dot{E_{\bm{k}}}-aB_{\bm{k}}\right)\right] .
\eea 

The metric perturbations that feature in the right-hand side of \Eq{eq:pertubations:general} satisfy the Einstein field equations, and in particular the energy and momentum constraints 
\begin{align}
\label{eq:energyconstraint:arbgauge} 3H\left(\dot{\psi_{\bm{k}}}+HA_{\bm{k}}\right) + \frac{k^2}{a^2}\left[\psi_{\bm{k}} + H\left(a^2\dot{E_{\bm{k}}}-aB_{\bm{k}}\right)\right] &= -\frac{1}{2\Mp^2}\left[ \dot{\phi}\left(\dot{\delta\phi_{\bm{k}}}-\dot{\phi}A_{\bm{k}}\right)+V_{,\phi}\delta\phi_{\bm{k}}\right], \\
\label{eq:momentumconstraint:arbgauge} \dot\psi_{\bm{k}} + H A_{\bm{k}} &= \frac{\dot\phi}{2\Mp^2}  \delta\phi_{\bm{k}} \, .
\end{align}
Introducing the Sasaki--Mukhanov variable~\cite{Sasaki:1986hm, Mukhanov:1988jd}
\bea \label{eq:def:Q}
Q_{\bm{k}} = \delta\phi_{\bm{k}} + \frac{\dot{\phi}}{H}\psi_{\bm{k}} \, ,
\eea
and using \Eqs{eq:energyconstraint:arbgauge} and \eqref{eq:momentumconstraint:arbgauge} to eliminate the metric perturbations, \Eq{eq:pertubations:general} can be rewritten as 
\bea \label{eq:pertubations}
\ddot{Q}_{\bm{k}} + 3H\dot{Q}_{\bm{k}} + \left[ \frac{k^2}{a^2} + V_{,\phi\phi} - \frac{1}{a^3\Mp^2}\frac{\dd}{\dd t}\left( \frac{a^3}{H}\dot{\phi}^2 \right) \right]Q_{\bm{k}} = 0 \, .
\eea 
We now compare this equation with the one coming from perturbing the background equations.

\subsection{Perturbed background equations}

In order to easily relate the field fluctuation $\delta\phi$ to the Sasaki--Mukhanov variable, one usually chooses to work in the spatially-flat gauge where $\psi=0$, and hence $Q=\delta\phi$ according to \Eq{eq:def:Q}. In this section, we will show how to perturb the background equations in that gauge (see \Sec{sec:PertBackEOM:SFG}), but also in the uniform-$N$ gauge that is used in stochastic inflation (see \Sec{sec:PertBackEOM:UNG}). 

Let us perturb the quantities appearing in \Eq{eq:kleingordon}, according to 
\bea \label{eq:perturbkleingordon}
&\phi \to \phi + \delta\phi \, , &\dd t \to (1+A) \dd t \, ,
\eea 
where $1+A$ is the lapse function introduced in \Eq{eq:perturbedlineelement:FLRW}. Let us stress that the lapse function needs to be perturbed, otherwise one is implicitly working in a synchronous gauge (where $A=0$), which in general differs from the spatially-flat and uniform-$N$ gauges, and this leads to inconsistencies~\cite{Cruces:2018cvq}. Inserting \Eq{eq:perturbkleingordon} into \Eq{eq:kleingordon} gives rise to
\bea \label{eq:perturbedKG}
\ddot{\delta\phi} + \left(3H + \frac{\dot{\phi}^2}{2\Mp^2H}\right)\dot{\delta\phi} \: + &\left(\frac{\dot{\phi}}{2\Mp^2H}V_{,\phi} + V_{,\phi\phi}\right)\delta\phi
\\
- &\dot{\phi}\dot{A} - \left(2\ddot{\phi}+3H\dot{\phi}+ \frac{\dot{\phi}^3}{2\Mp^2H}\right)A = 0 \, ,
\eea 
where we have also used 
\bea \label{eq:deltaH} 
\delta H = \frac{V_{,\phi}\delta\phi + \dot{\phi}\dot{\delta\phi} - \dot{\phi}^2A}{6\Mp^2 H}  
\eea 
that comes from perturbing the Friedmann equation \eqref{eq:friedmann} under \Eq{eq:perturbkleingordon}.

\subsubsection{Spatially-flat gauge}
\label{sec:PertBackEOM:SFG}

In the spatially-flat gauge, the lapse function can readily be rewritten in terms of the field perturbation by imposing the momentum constraint~\eqref{eq:momentumconstraint:arbgauge}, which simplifies to 
\bea
\label{eq:flatA}
A = \frac{\dot{\phi}}{2\Mp^2 H} \delta \phi \,.
\eea
Substituting this relation into \Eq{eq:perturbedKG} gives rise to 
\bea \label{eq:perturbedKG:simplified}
\ddot{\delta\phi} + 3H\dot{\delta\phi} + \left[ V_{,\phi\phi} - \frac{1}{\Mp^2a^3}\frac{\dd}{\dd t}\left(\frac{a^3}{H}\dot{\phi}^2\right) \right] \delta\phi = 0 \, .
\eea 
Comparing \Eq{eq:perturbedKG:simplified}, obtained from the perturbed background equations, with \Eq{eq:pertubations}, obtained in linear perturbation theory in the spatially-flat gauge where $Q=\delta\phi$, we see that the two are consistent in the super-Hubble limit where $k\ll a H$. 

It is important to note that the local proper time in each patch is perturbed with respect to the cosmic time, $t$, in the background in the presence of a non-zero lapse perturbation, $A$. As can be seen from \Eq{eq:flatA} the perturbation $A$ vanishes in the spatially-flat gauge in the slow-roll limit, $\dot\phi\to0$, and the local proper time in this limit coincides with the background cosmic time. Beyond slow roll one must consistently account for local variations in the proper time interval in different patches if one wants to relate the separate universe equations to the perturbation equations written in terms of a global (background) cosmic time. This will be the aim of \Sec{sec:uniformexpansion}.

\subsubsection{Uniform-$N$ gauge}
\label{sec:PertBackEOM:UNG}
Let us introduce the expansion rate of $t=$constant hypersurfaces
\bea
\theta={n^\mu}_{;\mu}\, ,
\eea
where $n^\mu$ is the unit time-like vector, orthogonal to the constant-time hypersurfaces.
It is related to the metric perturbations in \Eq{eq:perturbedlineelement:FLRW} according to~\cite{Malik:2008im} 
\bea \label{eq:def:expansion}
\theta = \frac{3}{a}\left( \mathcal{H} - \mathcal{H}A - \psi' + \frac{1}{3}\nabla^2\sigma \right) \, ,
\eea 
where $\mathcal{H} = a'/a$ is the conformal Hubble parameter, a prime is a derivative with respect to conformal time $\eta$ defined through $\dd t = a \dd \eta$, and $\sigma = E'-B$ is the shear potential.
From the perturbed expansion rate $\theta$, one can define a perturbed integrated expansion up to first order in the metric perturbations
\bea 
\tilde{N} &= \frac{1}{3}\int \theta (1+A) \dd t 
= N - \psi + \frac{1}{3}\nabla^2\int\sigma\dd\eta \, .
\eea 
The last term in the right-hand side can be re-written in terms of $\E\equiv \int \sigma\dd\eta$, which corresponds to $E$ in the hypersurface-orthogonal threading where $B=0$. From now on, we work in such a spatial threading. This gives rise to
\bea  \label{eq:def:deltaN}
\delta N = -\psi + \frac{1}{3}\nabla^2\E \, ,
\eea 
i.e., the perturbation of the trace of the spatial metric on constant-time hypersurfaces.
Note, in particular, that in the spatially-flat gauge where $\psi=B=0$, we have $\delta N|_{\psi=0} = \frac{1}{3}\nabla^2\E |_{\psi=0}$. 

The uniform-$N$ gauge used in the Langevin equations \eqref{eq:conjmomentum:langevin} and \eqref{eq:KG:efolds:langevin} is defined by keeping the integrated expansion unperturbed across all patches of the universe, \ie $\delta N = 0$. From \Eq{eq:def:deltaN}, this imposes a direct relationship between $\psi$ and $E$, namely $\psi = \frac{1}{3}\nabla^2E_B$.
In the uniform-$N$ gauge, we note that the perturbation equation \eqref{eq:pertubations:general} can be written as
\bea 
\label{eq:pert:uniformN}
\ddot{\delta\phi_{\bm{k}}} + 3H\dot{\delta\phi_{\bm{k}}} + \left( \frac{k^2}{a^2} + V_{,\phi\phi} \right)\delta\phi_{\bm{k}} &= \dot{\phi}\dot{A_{\bm{k}}} - 2V_{,\phi}A_{\bm{k}}  \\
&= \dot{\phi}\dot{A_{\bm{k}}} + \left(2\ddot{\phi} + 6H\dot{\phi}\right)A_{\bm{k}} \, .
\eea
This can be recast in a form similar to the perturbed background equation~\eqref{eq:perturbedKG}, namely
\bea 
\label{eq:KGuniformN}
\ddot{\delta\phi_{\bm{k}}} + \left(3H + \frac{\dot{\phi}^2}{2\Mp^2H}\right)\dot{\delta\phi_{\bm{k}}} &+ \left( \frac{\dot{\phi}}{2\Mp^2H}V_{,\phi} +V_{,\phi\phi}\right)\delta\phi_{\bm{k}}
 \\
& - \dot{\phi}\dot{A_{\bm{k}}} - \left( 2\ddot{\phi}+3H\dot{\phi} + \frac{\dot{\phi}^3}{2\Mp^2H} \right)A_{\bm{k}} = \Delta_{\bm{k}}
\eea
where the difference between \Eqs{eq:perturbedKG} and \eqref{eq:pert:uniformN} is quantified as
\bea \label{eq:difference}
\Delta_{\bm{k}} = \frac{\dot{\phi}}{H}\left\lbrace \frac{1}{2\Mp^2}\left[\dot{\phi}\left(\dot{\delta\phi}_{\bm{k}} - \dot{\phi}\dot{A}_{\bm{k}}\right) + V_{,\phi}\delta\phi_{\bm{k}}\right] + 3H^2A_{\bm{k}}\right\rbrace - \frac{k^2}{a^2}\delta\phi_{\bm{k}} \, .
\eea 
If we now impose the energy constraint \eqref{eq:energyconstraint:arbgauge} in the uniform-$N$ gauge, and recalling that since we choose $B=0$, $\psi = \frac{1}{3}\nabla^2E$, one can show that
\bea \label{eq:UN:diff}
\Delta_{\bm{k}} = - \frac{k^2}{a^2}\left( \delta\phi_{\bm{k}} + \frac{\dot\phi}{H}\psi_{\bm{k}} \right) = - \frac{k^2}{a^2}Q_{\bm{k}} \, ,
\eea 
see \Eq{eq:def:Q}.
Hence, since we neglect $k^2/a^2$ terms in the large-scale limit, the perturbation equations and the perturbed background equations become identical on large scales. We conclude that the separate universe approach, describing the evolution of long-wavelength perturbations about an FLRW background in terms of locally FLRW patches, is valid in both the spatially-flat and uniform-$N$ gauges. This result does not rely on slow roll; we only require that we can neglect gradient terms on super-Hubble scales. 
\subsection{Arbitrary gauge}
Let us finally see how the above arguments can be formulated without fixing a gauge. 
It is instructive to collect together metric perturbation terms in the Klein-Gordon equation from the full linear perturbation theory, \Eq{eq:pertubations:general}, which describe the perturbation of the local expansion rate \eqref{eq:def:expansion}
\bea
\delta\theta_{\bm{k}} = - 3\dot\psi_{\bm{k}} - \frac{k^2}{a^2}\left(a^2\dot{E_{\bm{k}}}-aB_{\bm{k}}\right) - 3HA_{\bm{k}} \,.
\eea
Re-writing the perturbed Klein-Gordon equation \eqref{eq:pertubations:general} in terms of $\delta\theta_{\bm{k}}$ we obtain
\bea
\label{eq:pertubations:general2}
\ddot{\delta\phi_{\bm{k}}} + 3H\dot{\delta\phi_{\bm{k}}} + \left(\frac{k^2}{a^2}+V_{,\phi\phi}\right)\delta\phi_{\bm{k}} 
= \left(2\ddot\phi+3H\dot\phi\right)A_{\bm{k}} + \dot{\phi} \dot{A_{\bm{k}}} 
- \dot{\phi}
\delta\theta_{\bm{k}}
\,.
\eea 
Finally, combining \Eq{eq:pertubations:general2} with the background equation \eqref{eq:kleingordon} and rewriting the time derivatives in terms of the local proper time rather than the coordinate time, $\partial/\partial\tau\equiv(1-A)\partial/\partial t$, one obtains
\bea
\frac{\partial^2}{\partial\tau^2}(\phi+\delta\phi) + \theta\frac{\partial}{\partial\tau}(\phi+\delta\phi) + V_{,\phi}(\phi+\delta\phi) = \frac{\nabla^2}{a^2} (\delta\phi) \,.
\eea
Thus we see that the perturbed Klein-Gordon equation~\eqref{eq:pertubations:general} from cosmological perturbation theory in an arbitrary gauge has exactly the same form, up to first order in the inhomogeneous field and metric perturbations and up to spatial gradient terms of order $\nabla^2\delta\phi$, as the Klein-Gordon equation for a homogeneous scalar field in an FLRW cosmology, \Eq{eq:kleingordon}, where we identify the local proper time, $\tau$, with the coordinate time, $t$, in an FLRW cosmology and the local expansion rate, $\theta/3$, with the Hubble rate, $H$, in an FLRW cosmology.
However to relate these local quantities to a global background coordinate system we need to fix a gauge. This cannot be determined by the local FLRW equations but requires to use additional constraint equations from the cosmological perturbation theory, as demonstrated in the preceding sub-sections for the spatially-flat and uniform-$N$ gauges.

\section{Gauge corrections to the noise} 
\label{sec:uniformexpansion}

In the previous section, it was explained that the field fluctuations, which determine the noise correlators through \Eq{eq:noise:correlators}, are usually calculated in the spatially-flat gauge, where the field perturbations coincide with the Sasaki--Mukhanov variable. However, we have seen that the local time in the spatially-flat gauge is in general perturbed with respect to the global time. As stressed in \Sec{sec:stochastic}, the Langevin equations (\ref{eq:conjmomentum:langevin}) and~(\ref{eq:KG:efolds:langevin}) are written in terms of the number of \efolds, i.e., the integrated expansion, $N$, is used as a time variable. If we are to use the number of \efolds\ as a local time coordinate and also as a global coordinate, relating the stochastic distribution of field values in many different patches at a given time, then we need to work in the uniform-$N$ gauge.
Thus one needs to gauge transform the field fluctuations calculated in the spatially-flat gauge to the uniform-$N$ gauge before evaluating \Eq{eq:noise:correlators}, and in this section, we explain how this can be done.
\subsection{Gauge transformations}
Let us denote quantities in the uniform-$N$ gauge with a tilde, \ie $\widetilde{\delta N}=0$. The transformations from the spatially-flat to the uniform-$N$ gauge can be written by means of a gauge transformation parameter $\alpha$ (that will be determined below), according to~\cite{Malik:2008im}
\begin{align} 
\label{eq:transform:phi} \delta\phi &\to \widetilde{\delta\phi} = \delta\phi + \phi'\alpha\, , \\
\label{eq:transform:psi}\psi &\to \widetilde{\psi} = \psi - \mathcal{H}\alpha\, , \\
\E &\to \widetilde{\E} = \E + \int\alpha\dd\eta \, .
\end{align}
Combining these transformation rules with \Eq{eq:def:deltaN}, the perturbed integrated expansion transforms as 
\bea \label{eq:transform:deltaN}
\delta N \to \widetilde{\delta N} = \delta N + \mathcal{H}\alpha + \frac{1}{3}\nabla^2\int\alpha \dd \eta \, .
\eea 
By definition, $\widetilde{\delta N}=0$, so one is lead to
\bea \label{eq:integraleq:alpha}
\delta N\Big|_{\psi=0} + \mathcal{H}\alpha  + \frac{1}{3}\nabla^2\int\alpha\dd\eta = 0 \, .
\eea
Taking the derivative of this expression with respect to conformal time, one obtains a differential equation for the gauge transformation parameter $\alpha$, namely
\bea \label{eq:alpha:diff}
3\mathcal{H}\alpha' + \left(3\mathcal{H}'+ \nabla^2\right)\alpha = S
 \, ,
\eea
where the source term reads
\bea
S = - 3 \delta N'\Big|_{\psi=0} = - \nabla^2 \sigma \Big|_{\psi=0} \,.
\eea

Two remarks are then in order. First, the source standing on the right-hand side of \Eq{eq:alpha:diff} remains to be calculated. In \Sec{sec:Nad:press}, we will show that for a scalar field it is related to the non-adiabatic pressure perturbation, and we will explain how it can be calculated. In \Sec{sec:source}, we will provide the general solution to \Eq{eq:alpha:diff}. Second, once $\alpha$ is determined, the field fluctuations in the uniform-$N$ gauge can be obtained from those in the spatially-flat gauge via \Eq{eq:transform:phi}. The noise correlators~(\ref{eq:noise:correlators}) also involve the fluctuation in the conjugate momentum, so this needs to be transformed into the uniform-$N$ gauge as well. However, precisely since $N$ is unperturbed in the uniform-$N$ gauge, one simply has
\bea 
\label{eq:GaugeTransf:pi}
\widetilde{\delta \pi} & = \frac{\dd \widetilde{\delta\phi}}{\dd N} \, ,
\eea 
and $\widetilde{\delta \pi}$ can be inferred from $\widetilde{\delta\phi}$ by a straightforward time derivative.
\subsection{Non-adiabatic pressure perturbation}
\label{sec:Nad:press}
Let us now show that the source function, $S(\eta)$ of \Eq{eq:alpha:diff}, 
coincides with the non-adiabatic pressure perturbation for a scalar field. This will prove that if inflation proceeds along a phase-space attractor (such as slow roll), where non-adiabatic pressure perturbations vanish, the source function vanishes as well; in this case \Eq{eq:alpha:diff} is solved by $\alpha=0$, and there are no gauge corrections.

Let us start by recalling the expressions for the energy constraint in an arbitrary gauge~\cite{Malik:2008im}
\begin{align} \label{eq:energyconstraint}
3\mathcal{H}\left(\psi' + \mathcal{H}A\right) - \nabla^2\left(\psi + \mathcal{H}\sigma\right)
&= 
-\frac{a^2}{2\Mp^2}\delta\rho \, . 
\end{align}
Combining this with the momentum constraint \eqref{eq:momentumconstraint:arbgauge} gives 
\bea \label{eq:deltarhodeltaphi}
\nabla^2(\psi+\mathcal{H}\sigma) &= \frac{a^2}{2\Mp^2}
\delta\rho_{\mathrm{com}}
\, ,
\eea
where the comoving density perturbation for a scalar field is given by 
\bea \label{eq:def:comdensity}
\delta\rho_{\mathrm{com}} = \delta\rho - 
\frac{\rho'}{\phi'}\delta\phi \, .
\eea
This in turn can be related to the non-adiabatic pressure perturbation~\cite{Malik:2008im}
\bea \label{eq:def:deltaPnad}
\delta P_{\mathrm{nad}} = -\frac{2a^2}{3\mathcal{H}\phi'}V_{,\phi}\delta\rho_{\mathrm{com}} \, .
\eea
In particular, in the spatially-flat gauge where $\psi=0$, \Eq{eq:deltarhodeltaphi} becomes
\bea \label{eq:energyconstraint:Pnad}
\mathcal{H}\nabla^2\sigma|_{\psi=0}
= -\frac{3\mathcal{H}\phi'}{4\Mp^2V_{,\phi}} \delta P_{\mathrm{nad}} \, .
\eea
Thus the source term $S$ on the right-hand side of \Eq{eq:alpha:diff} reads
\bea
S = \frac{3\phi'}{4\Mp^2V_{,\phi}} \delta P_{\mathrm{nad}}\,,
\eea  
and it vanishes if the non-adiabatic pressure perturbation is zero, which is the case whenever inflation proceeds along a phase-space attractor, $\phi'=\phi'(\phi)$, such as during slow roll.

In order to find a general expression for $S(\eta)$, one can use the (arbitrary gauge) expression for $\delta\rho$ for a scalar field~\cite{Malik:2008im},
\bea \label{eq:constraint:rho}
\delta\rho = \frac{\phi'\delta\phi' - \phi'^2A}{a^2} + V_{,\phi}\delta\phi \, ,
\eea
and combine it with \Eq{eq:deltarhodeltaphi} to obtain
\bea 
\nabla^2(\psi+\mathcal{H}\sigma) &= \frac{a^2}{2\Mp^2}\left[\left(3H\dot{\phi} + V_{,\phi}\right)\delta\phi + \dot{\phi}\delta\dot{\phi} - \dot{\phi}^2A\right] \, .
\eea 
Hence, in terms of the field fluctuations in the spatially-flat gauge and using \Eq{eq:flatA} for the perturbed lapse function, one finds 
\bea
S = -\frac{1}{2\Mp^2\mathcal{H}}\left[\left(3\mathcal{H}\phi' + a^{2}V_{,\phi} - \frac{\phi'^3}{2\Mp^2\mathcal{H}} \right)Q + \phi'Q'\right]\, .
\eea
Introducing the second slow-roll parameter $\epsilon_2\equiv \dd\ln\epsilon_1/\dd N$, the source function can be rewritten in the simpler form
\bea \label{eq:sourcefunction:general}
S = \frac{Q\sqrt{2\epsilon_1}}{2\Mp} \mathrm{sign}(\dot{\phi}) \left(  \calH\frac{\epsilon_{2}}{2} - \frac{Q'}{Q}\right) \, .
\eea
\subsection{General solution}
\label{sec:source}
When written in Fourier space, the differential equation~\eqref{eq:alpha:diff} for $\alpha_k$ has the general solution
\bea \label{eq:alphaintegral1:general}
\alpha_k &= \frac{1}{3\mathcal{H}}\int^{\eta}_{\eta_0} S_k(\eta')\exp\left[ \frac{k^2}{3}\int^{\eta}_{\eta'}\frac{\dd\eta''}{\mathcal{H}(\eta'')} \right]\dd\eta' \, .
\eea 
In this expression, $\eta_0$ is an integration constant that defines the slicing relative to which the expansion is measured. In what follows, we will consider situations in which an attractor is reached at late times. Since, in such a regime, the gauge correction vanishes (given that the non-adiabatic pressure perturbation does), we will take $\eta_0$ in the asymptotic future, \ie $\eta_0 = 0^-$.

Finally, in \Eq{eq:sourcefunction:general} the Sasaki--Mukhanov variable, $Q$, needs to be determined, which can be done by solving the Sasaki--Mukhanov equation
\bea \label{eq:MSequn}
v_k'' + \left(k^2 - \frac{z''}{z}\right)v_k = 0 \, ,
\eea
where $v_k = aQ_k$ and $z \equiv a\sqrt{2\epsilon_{1}}\Mp $. One can show that, in full generality $z''/z=\calH^2(2-\epsilon_1+3\epsilon_2/2-\epsilon_1\epsilon_2/2+\epsilon_2^2/4+\epsilon_2\epsilon_3/2)$, where we have introduced the third slow-roll parameter $\epsilon_3\equiv \dd\ln\epsilon_2/\dd\ln N$. For future use however, instead of working with the second and third slow-roll parameters, it will be more convenient to work with the field acceleration parameter
\bea
\label{eq:def:f} f&=-\frac{\ddot{\phi}}{3H\dot{\phi}} = 1+\frac{1}{3H\dot{\phi}}V_{,\phi}
\eea
and the dimensionless mass parameter
\bea
\label{mu:def}
\mu &= \frac{V_{,\phi\phi}}{3H^2} \, ,
\eea
in terms of which
\bea \label{eq:z''overz:general}
\frac{z''}{z} &= \calH^2 \left( 2 + 5\epsilon_{1} - 3\mu - 12f\epsilon_{1} + 2\epsilon_{1}^2 \right) ,
\eea 
see  \App{appendix:MSequation}. 

The following three sections will be devoted to three case studies, for which \Eq{eq:MSequn} will be solved and \Eq{eq:alphaintegral1:general} will be evaluated in order to derive the gauge corrections to the stochastic noise correlators in the uniform-$N$ gauge with respect to those in the spatially-flat gauge. In all cases, we will find that at the order of the coarse-graining parameter at which the stochastic formalism is derived, these gauge corrections can be neglected.
\section{Case study 1: slow roll}
\label{sec:slowroll}
Let us first apply the programme sketched above to the case of slow-roll inflation. As argued before, the presence of a dynamical attractor in that case makes the non-adiabatic pressure perturbation vanish, hence we should not find any gauge correction to the field fluctuations in the uniform-$N$ gauge and thus to the correlators for the noise. This is therefore a consistency check of our formalism. 

At leading order in slow roll, the slow-roll parameters can simply be evaluated at the Hubble-crossing time $\eta_{*}\simeq -1/k$, since their time dependence is slow-roll suppressed, \ie $\epsilon_{1} = \epsilon_{1*} + \mathcal{O}(\epsilon^2)$, etc. At that order, \Eq{eq:MSequn} is solved according to 
\bea \label{eq:MS:SRsolution}
v_k = \frac{\sqrt{-\pi\eta}}{2}\mathrm{H}_{\nu}^{(2)}\left(-k\eta\right) = aQ_k \, ,
\eea
where $\mathrm{H}_{\nu}^{(2)}$ is the Hankel function of the second kind and $\nu \equiv 3/2+\epsilon_{1*}+\epsilon_{2*}/2$, see \App{appendix:solving:MSequation}. Since the coarse-graining parameter is such that $\sigma\ll 1$, the mode functions in \Eq{eq:noise:correlators} need to be evaluated in the super-Hubble regime, \ie when $-k\eta \ll 1$. One can therefore make use of the asymptotic behaviour
\bea \label{eq:Hankel2:largescale}
H_{\nu}^{(2)}(-k\eta) \simeq \frac{i\Gamma(\nu)}{\pi}\left(\frac{2}{-k\eta}\right)^{\nu}\left[1 + \frac{1}{4(\nu-1)}\left(-k\eta\right)^{2} +\order{k^4\eta^4}\right] \, .
\eea
On the other hand, at first order in slow roll, the scale factor can also be expanded, and one finds
\bea \label{eq:scalefactor:slowroll}
a = -\frac{1}{H_{*}\eta}\left[1 + \epsilon_{1*} - \epsilon_{1*}\ln{\left(\frac{\eta}{\eta*}\right)}+\mathcal{O}(\epsilon^2)\right] \, ,
\eea
where we have used the expression $\mathcal{H}\simeq -\frac{1}{\eta}\left( 1 + \epsilon_{1*} \right)$ derived in  \App{sec:z''/z:SR}. Combining the two previous equations then leads to
\bea \label{eq:Q'overQ:SR}
\frac{Q'_k}{Q_k} \simeq \frac{\frac{3}{2}+\epsilon_{1*}-\nu}{\eta} + \frac{\frac{7}{2}-\nu}{4(\nu-1)}k^2\eta \, ,
\eea
which is valid at next-to-leading order both in the slow-roll parameters and in $k\eta$. With the expression given above for $\nu$, one can see that $Q'_k/Q_k\simeq -\epsilon_2/(2\eta)$ at leading order in $k\eta$. Since, at leading order, $\calH\simeq -1/\eta$, the two terms in the right-hand side of \Eq{eq:sourcefunction:general} exactly cancel, and the source function $S_k$ vanishes. This confirms that the gauge corrections are indeed suppressed in that case.

In fact, the first contribution to the gauge correction comes from the decaying mode, and for completeness we now derive its value. Plugging the previous expressions into \Eq{eq:sourcefunction:general} leads to
\bea \label{eq:sourcefunction:SR}
S_k &= \frac{i}{2}\frac{H_*}{\Mp}\sqrt{k\epsilon_{1*} }\eta\left(-k\eta\right)^{-\epsilon_{1*}-\frac{\epsilon_{2*}}{2}}  \mathrm{sign}\left(\dot{\phi}\right) \, .
\eea
One can then insert \Eq{eq:sourcefunction:SR}, along with $\mathcal{H} = -(1+\epsilon_{1*})/\eta$, into \Eq{eq:alphaintegral1:general}, and derive the gauge transformation parameter from the spatially-flat gauge to the uniform-$N$ gauge in the large-scale and slow-roll limit, 
\bea \label{eq:alphaLO:SR}
\alpha_k = \frac{iH_{*}\sqrt{\epsilon_{1*}}}{12\Mp}k^{-\frac{5}{2}}\left(-k\eta\right)^{3-\epsilon_{1*}-\frac{\epsilon_{2*}}{2}}  \mathrm{sign}\left(\dot{\phi}\right) \, .
\eea
In the uniform-$N$ gauge, according to \Eq{eq:transform:phi}, the field fluctuation thus reads
\bea 
\label{eq:GaugeCorr:SR}
\widetilde{\delta\phi}_k &= Q_k\left[ 1 - \frac{\epsilon_{1*}}{6}\left(-k\eta\right)^2 \right] ,
\eea
and its deviation from $Q$ is therefore both slow-roll suppressed and controlled by the amplitude of the decaying mode. Since it needs to be evaluated at the coarse-graining scale $k_\sigma = \sigma a H$ in \Eq{eq:noise:correlators}, the relative gauge correction to the correlations of the noises scales as $\epsilon_1\sigma^2$, which can be neglected since the stochastic formalism assumes $\sigma\rightarrow 0$.
\section{Case study 2: ultra-slow roll}
\label{sec:USR}
Let us now consider the case of ultra-slow-roll (USR) inflation~\cite{Inoue:2001zt, Kinney:2005vj, Pattison:2018bct}, where the dynamics of $\phi$ is friction dominated and the gradient of the potential can be neglected in the Klein--Gordon equation~\eqref{eq:kleingordon}, which becomes
\bea 
\ddot{\phi} + 3H\dot{\phi} \simeq 0 \, .
\eea
This gives rise to $\dot{\phi}_{\mathrm{USR}} \propto \ee^{-3N}$, hence $\dot{\phi}=\dot{\phi}_\uin+3H(\phi_\uin-\phi)$. The phase-space trajectory thus carries a dependence on initial conditions that is not present in slow roll, which explains why ultra-slow roll is not a dynamical attractor while slow roll is. We therefore expect the non-adiabatic pressure perturbation not to vanish in ultra-slow roll, which may lead to some non-trivial gauge corrections. In ultra-slow roll,  the field acceleration parameter $f$ introduced in \Eq{eq:def:f} is close to one (while it is close to zero in slow roll), so $\delta\equiv 1-f$ quantifies how deep in the ultra-slow-roll regime one is. In the limit where $\delta=0$,  $\dot{\phi}_{\mathrm{USR}} \propto \ee^{-3N}$ gives rise to $\epsilon_1^{\mathrm{USR}}\propto \ee^{-6N}/H^2$, hence
\bea \label{eq:eps:USR}
\epsilon_n^{\mathrm{USR}} = \begin{cases}
-6 + 2\epsilon_{1} &\text{if $n$ is even}\\
2\epsilon_{1} &\text{if $n>1$ is odd} \, .
\end{cases}
\eea 
The even slow-roll parameters are therefore large in ultra-slow roll. When $\delta$ does not strictly vanish, these expressions can be corrected, and for the second and the third slow-roll parameters, one finds
\begin{align}
\label{eq:eps2:USR}
\epsilon_{2}&= -6  + 2\epsilon_{1}+6\delta \,, \\
\epsilon_{3}&= 2\epsilon_{1} - \frac{\dd \delta}{\dd N}\frac{6}{6 - 2\epsilon_{1} - 6\delta} \, ,
\end{align}
which are exact formulas. One can then calculate
\bea \label{eq:ddeltadt:USR}
\frac{\dd \delta}{\dd N} &=  -\mu + 3\delta - 3\delta^2 + \delta\epsilon_{1} \, ,
\eea 
where $\mu$ is the dimensionless mass parameter defined in \Eq{mu:def}. For small $\delta$ and $\epsilon_{1}$, one then has
\bea 
\epsilon_{3}^{\mathrm{USR}} \simeq 2\epsilon_{1} + \mu - 3\delta + \mu\left( \frac{2\epsilon_{1}+6\delta}{6} \right) \, .
\eea 
There is no reason, \textit{a priori}, that $\mu$ needs to be small, and hence these corrections can be large for models with $V_{,\phi\phi}\neq 0$.
Note also that \Eq{eq:ddeltadt:USR} provides a criterion for the stability of ultra-slow roll, which is stable when the right-hand side of this equation is negative, in agreement with the results of \Ref{Pattison:2018bct}. 

Let us now derive the gauge corrections in ultra-slow roll. We perform a calculation at leading order in $\epsilon_1$, $\delta$ and $\mu$, but in \App{app:usr:epscorrections} the calculation is extended to next-to-leading order in $\epsilon_1$, and it is shown that the result derived below is indeed valid. At leading order, one simply has $z''/z\simeq 2 \calH^2$, hence \Eq{eq:MSequn} is solved according to
\bea \label{eq:v:nu=3/2}
v_k &= \frac{1}{\sqrt{2k}}\ee^{-ik\eta}\left( 1 - \frac{i}{k\eta}\right) \, .
\eea 
Since $a = -1/(H_{*}\eta)$ at leading order, this gives rise to
\bea \label{eq:Q'overQ:USR:nu=3over2}
\frac{Q'_k}{Q_k} = \frac{-ik^2\eta}{k\eta-i} \, ,
\eea 
and the source function~\eqref{eq:sourcefunction:general} reads
\bea \label{eq:source:USR:nu=3/2}
S_k &= \frac{H_{*}}{2\Mp}\sqrt{\frac{\epsilon_{1}}{k}}\ee^{-ik\eta}\left( 3 - \frac{3i}{k\eta} + ik\eta \right) \mathrm{sign}\left(\dot{\phi}\right) \, .
\eea
Since $\epsilon_{1}\simeq \epsilon_{1*}(a/a_*)^{-6}$, the gauge transformation parameter $\alpha$ can be obtained from \Eq{eq:alphaintegral1:general} and is given by
\bea \label{eq:alpha:USR:nu=3/2}
\alpha_k = \frac{iH_{*}\sqrt{\epsilon_{1*}}}{6\Mp}{k^{-\frac{5}{2}}}(k\eta)^4\mathrm{sign}\left(\dot{\phi}\right) \left[  1+ \mathcal{O}(k\eta)^2\right] \, .
\eea
Comparing this expression with \Eq{eq:alphaLO:SR}, one can see that the gauge correction decays even faster than in the slow-roll regime, hence is even more suppressed. This is because, although slow roll is a dynamical attractor while ultra-slow roll is not, the field velocity (hence the conjugate momentum) decays very quickly in ultra-slow roll, and this also damps away one of the two dynamical degrees of freedom. Finally, the gauge transformation \eqref{eq:transform:phi} gives rise to
\bea \label{eq:deltaphi:USRtransform}
\widetilde{\delta\phi_k} &= Q_k\left[ 1 + \frac{\epsilon_{1*}}{3}\left(-k\eta\right)^6\right]\, .
\eea
The relative corrections to the noises correlators scale as $\epsilon_{1*}\sigma^6$ and can therefore be neglected, even more accurately than in slow roll. 
\section{Case study 3: Starobinsky model}
\label{sec:starobinsky}
In the two previous sections, we have shown that the gauge corrections to the noise correlators are negligible both in slow-roll and in ultra-slow-roll inflation. In this section, we consider a model that interpolates between these two limits, namely the Starobinsky model~\cite{Starobinsky:1992ts}. This allows us to study a regime that is neither slow roll nor ultra-slow roll, but for which the early-time (ultra-slow roll) and the late-time (slow roll) limits are under control. 

The Starobinsky model is based on a potential made up of two linear parts with different gradients defined by the dimensionless parameters $a_+\gg a_- > 0$:
\bea 
V(\phi) = \begin{cases} V_{0}\left(1+a_+\frac{\phi}{\Mp}\right) & \mathrm{for}\, \phi> 0 \\ 
V_{0}\left(1+a_-\frac{\phi}{\Mp}\right) & \mathrm{for}\, \phi< 0 
\end{cases}
\, .
\eea
In order to ensure both parts of the potential are able to support slow-roll inflation,
we require $a_{\pm}\ll 1$. 

The dynamics of the inflaton, as it evolves across the discontinuity in the potential gradient at $\phi=0$, can be split into three phases. The first phase, which we label $\mathrm{SR}_{+}$, is a slow-roll phase for $\phi>0$ and $\dot\phi<0$. When the inflaton crosses $\phi=0$, it then starts down the $\phi<0$ part of the potential with an initial velocity inherited from the first slow-roll phase $\mathrm{SR}_{+}$ that is much larger than the slow-roll velocity for $\phi<0$. The second phase thus starts in an ultra-slow-roll regime and is denoted USR. It corresponds to the field range $\phi_{\mathrm{USR}\to\mathrm{SR}}<\phi<0$. Finally, the inflaton relaxes back to slow roll for $\phi<\phi_{\mathrm{USR}\to\mathrm{SR}}$, and we call this third phase $\mathrm{SR}_{-}$. 

\begin{figure}
    \centering
    \includegraphics[width=0.485\textwidth]{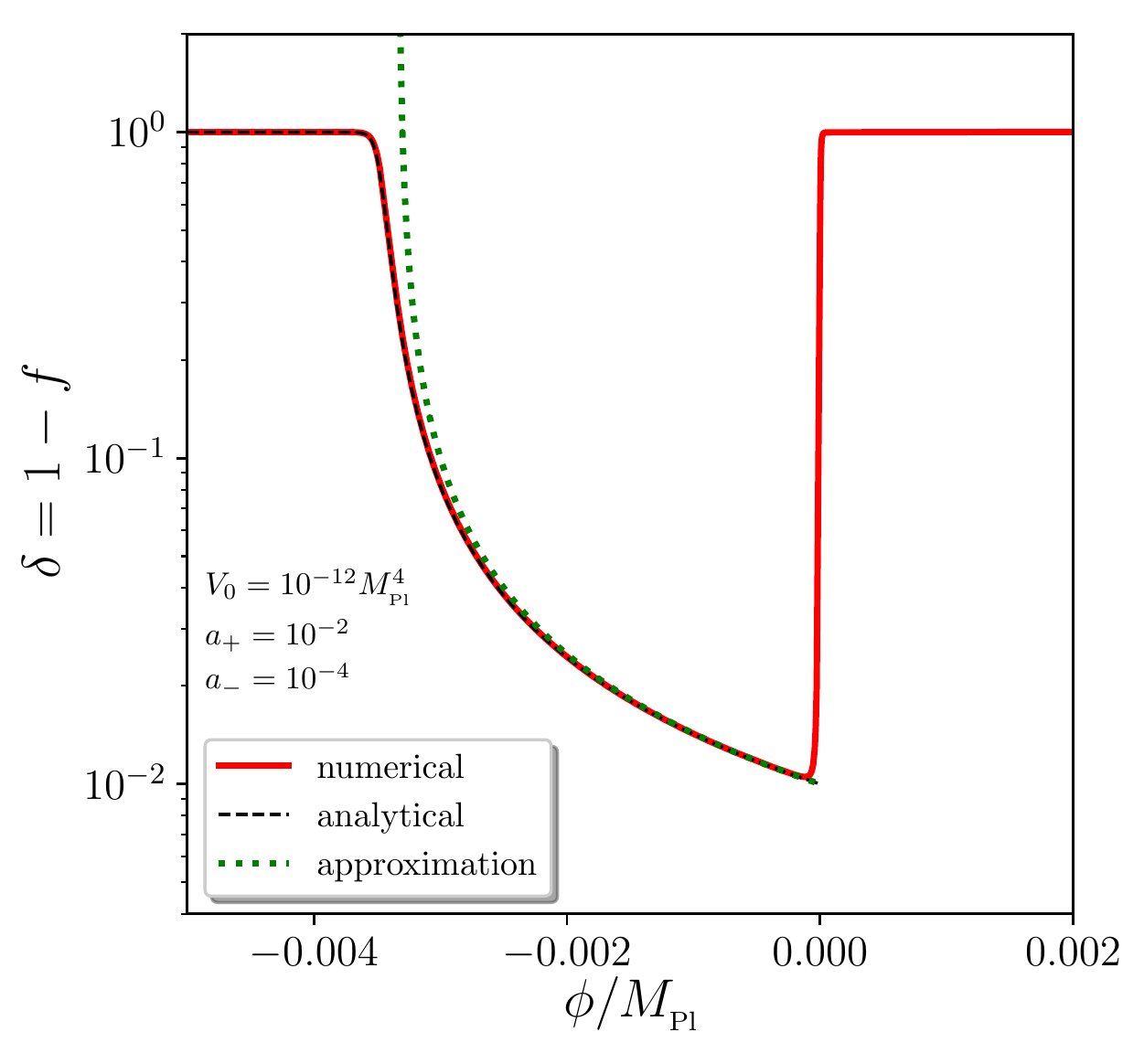}
    \includegraphics[width=0.50\textwidth]{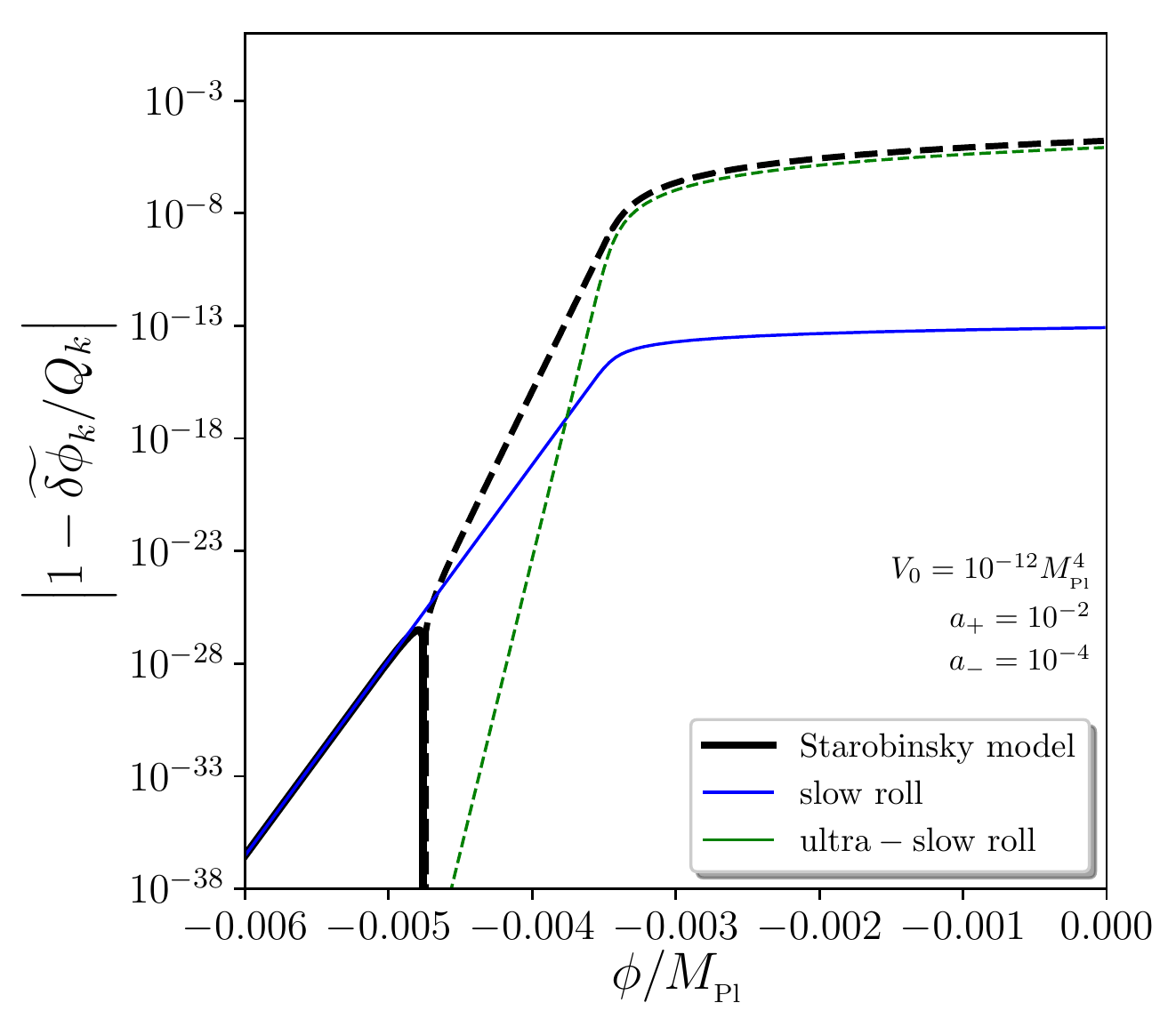}
    \caption{Left panel: field acceleration parameter, $f$ defined in \eqref{eq:def:f}, in the Starobinsky model. The red curve corresponds to a numerical integration of the Klein--Gordon equation~(\ref{eq:kleingordon}), the dashed black curve corresponds to the analytical solution~\eqref{eq:f:phi:starobinsky}, while the dotted green line stands for the approximation~(\ref{eq:f:phi:starobinsky:appr}). Right panel: fractional gauge correction to the field perturbation in the uniform-$N$ gauge in the Starobinsky model, for a mode such that $k/aH=10^{-2}$ at the transition time $t=0$. The black line corresponds to the full result~(\ref{eq:Staro:gaugeCorr}), the blue line stands for the slow-roll result~\eqref{eq:GaugeCorr:SR}, and the green line is the ultra-slow-roll result~\eqref{eq:deltaphi:USRtransform}. Solid lines are such that $1-\widetilde{\delta\phi}_k/Q_k>0$ and dashed lines are such that $1-\widetilde{\delta\phi}_k/Q_k<0$. }
    \label{fig:Staro}
\end{figure}
During the USR phase 
the Hubble parameter can be taken as approximately constant, $H\simeq H_0= \sqrt{V_0/(3\Mp^2)}$; the consistency of that assumption will be checked below. The Klein--Gordon equation~(\ref{eq:kleingordon}) then becomes $\ddot{\phi} + 3H_{0}\dot{\phi} + V_0 a_-/\Mp = 0$, and can be solved to give
\bea
\label{eq:phi:t:starobinsky}
\frac{\phi(t)}{\Mp}=\frac{a_+-a_-}{3}\left(\ee^{-3H_0t}-1\right)-a_- H_0 t\, ,
\eea
where we choose $t=0$ to denote the time when $\phi=0$, and the initial velocity is set such that its value at the transition point is given by its slow-roll counterpart in the $\phi>0$ branch of the potential, \ie $\dot{\phi}(\phi=0^-)=\dot{\phi}(\phi=0^+)=-H_0 a_+$. The acceleration parameter defined in \Eq{eq:def:f} is then given by
\bea 
\label{eq:Staro:f(t)}
f(t) &=  1 - \frac{a_-}{a_-+(a_+-a_-)\ee^{-3H_0t}} \, .
\eea 
At the transition time, it reads $f(t=0)= 1-\frac{a_-}{a_+}$, so if $a_-/a_+\ll 1$, $f\simeq 1$ and ultra-slow roll takes place. At late time, however, $f$ is damped so that the system relaxes back to a phase of slow-roll inflation. Note that the solution~\eqref{eq:phi:t:starobinsky} can be inverted, 
\bea 
\label{eq:Staro:traj:inverted}
H_0 t(\phi) = \frac{1}{3}\left(1-\frac{a_+}{a_-}\right) - \frac{\phi}{\Mp a_-} + \frac{1}{3}W_{0}\left[\frac{a_+-a_-}{a_-}\exp\left(
\frac{a_+}{a_-}-1+3 \frac{\phi}{\Mp a_-}\right)\right] \, ,
\eea 
where $W_{0}(x)$ is the $0$-branch of the Lambert function, which leads to the phase-space trajectory
\bea 
\label{eq:Staro:phidot(phi)}
\dot{\phi}(\phi) = -\frac{\Mp }{H} H_0^2 a_- \left\lbrace 1+W_{0}\left[\frac{a_+-a_-}{a_-}\exp\left(\frac{a_+}{a_-}-1+3 \frac{\phi}{\Mp a_-}\right)\right] \right\rbrace \, .
\eea 
In the denominator of the first term in the right-hand side, $H$ is left to vary~\cite{Martin:2011sn}, in such a way that at late time, \ie when $\phi$ goes to $-\infty$, one recovers the slow-roll result $\dot{\phi}=-\Mp H_0^2 a_-/H$. Plugging \Eq{eq:Staro:traj:inverted} into \Eq{eq:Staro:f(t)} also leads to 
\bea \label{eq:f:phi:starobinsky}
f(\phi) = 1 - \frac{1}{1+W_{0}\left[\frac{a_+-a_-}{a_-}\exp\left( 
\frac{a_+}{a_-}-1+3 \frac{\phi}{\Mp a_-}\right)\right]} \, ,
\eea 
which is shown in \Fig{fig:Staro} with the dashed black line and compared to a numerical solution of the Klein--Gordon equation displayed with the solid red line. One can check that $f$ starts from a value close to one at early time and approaches zero at late time. If one expands \Eq{eq:f:phi:starobinsky} around $\phi=0$, one obtains
\bea \label{eq:f:phi:starobinsky:appr}
f\simeq 1-\frac{a_-}{a_++3\frac{\phi}{\Mp}}\, ,
\eea
which matches Eq.~(4.3) of \Ref{Pattison:2018bct}. This approximation is also shown in \Fig{fig:Staro}, with the dotted green line. 

From \Eq{eq:f:phi:starobinsky}, the transition time between USR and $\mathrm{SR}_{-}$, defined as the time when $f=1/2$, is found to be
\bea 
t_{\mathrm{USR}\to\mathrm{SR}} = \frac{1}{3 H_0}\ln\left(\frac{a_+-a_-}{a_-}\right) \, ,
\eea 
which is consistent with Eq.~(4.5) of \Ref{Pattison:2018bct}. Making use of \Eq{eq:phi:t:starobinsky}, the field value at which this happens is given by
\bea 
\phi_{\mathrm{USR}\to\mathrm{SR}} &= -\frac{\Mp }{3}\left[ a_+ - 2 a_-+a_-\ln\left(\frac{a_+-a_-}{a_-}\right)\right]
 \simeq -\frac{a_+}{3}\Mp \, ,
\eea 
where the last expression is derived in the limit $a_-/a_+\ll 1$ and agrees with Eq.~(4.4) of \Ref{Pattison:2018bct}. This allows us to test the assumption made above that the potential, hence the Hubble parameter, does not vary much during the USR phase. The relative shift in the potential value between $\phi=0$ and $\phi_{\mathrm{USR}\to\mathrm{SR}}$ is indeed given by
\bea 
\frac{\Delta V}{V} = \frac{a_-(a_+-a_-)}{3} 
\ll 1 \, ,
\eea
which justifies the above assumption.

Let us now calculate the gauge transformation from the spatially-flat to uniform-$N$ gauge in this model. As explained above, combining \Eq{eq:Staro:phidot(phi)} and~(\ref{eq:phi:t:starobinsky}) leads to
\bea
\dot{\phi}(t) = \frac{H_0^2\Mp}{H}\left[ (a_--a_+)\ee^{-3H_{0}t} -  a_- \right] ,
\eea 
that allows us to both describe the USR and the $\mathrm{SR}_-$ phases, as well as the transition between the two. Making use of the relation $\epsilon_{1}=\dot{\phi}^2/(2\Mp^2H^2)$, one obtains
\bea 
\epsilon_{1}(t) &= \frac{1}{2}\left(\frac{H_0}{H}\right)^4\left[  a_- - (a_--a_+)\ee^{-3H_{0}t}\right]^2 \\
\epsilon_{2}(t) &= -\frac{6(a_--a_+)\ee^{-3H_0t}}{(a_--a_+)\ee^{-3H_0t}-a_-} + 4\epsilon_{1}(t) \, .
\eea 
One can check that, at late times, one recovers $\epsilon_2=4\epsilon_1$, which is indeed satisfied in slow roll for linear potentials, see \Eqs{eq:eps1:V} and~(\ref{eq:eps2:V}).

Since $\mu=0$ in this model, the fact that $\epsilon_1$ remains small implies that \Eq{eq:z''overz:general} is close to its de-Sitter limit. Moreover, one can check that, at early times, the term $Q'_k/Q_k$ in \Eq{eq:sourcefunction:general} provides a subdominant contribution, hence it is sufficient to evaluate $Q'_k/Q_k$ at late time and use the result of \Eq{eq:Q'overQ:SR}, $Q'_k/Q_k\simeq - \epsilon_{2*}/(2\eta)+k^2\eta = a_-^2/\eta+k^2\eta$. One then obtains
\bea 
S &= \frac{i  H_0}{\Mp}\frac{\mathrm{sign}\left(\dot{\phi}\right)}{(2k)^{\frac{3}{2}}\eta}\Bigg\{ 3(a_--a_+)\ee^{-3H_{0}t}\left(1+\frac{a_-^2}{3}\right)
 -  a_-^3  \\
&\hspace{1cm} 
 + \left[{a_-+(a_+-a_-)\ee^{-3H_{0}t}}\right]^3 - k^2\eta^2\left[ \left(a_--a_+\right)\ee^{-3H_{0}t} - a_-\right]\Bigg\} \, .
\eea 
From \Eq{eq:alphaintegral1:general}, we then find the gauge transformation parameter to be
\bea \label{eq:gaugetransformation:starobinsky}
\alpha &\simeq \frac{-i\eta H_0}{3(2k)^{\frac{3}{2}}\Mp}\mathrm{sign}\left(\dot{\phi}\right)
{\Bigg[}  
\frac{(k\eta)^2}{2}a_- 
+ \left(a_--a_+\right)\ee^{-3H_0t}\left(1+\frac{a_-^3}{2}\right)   \\
&\hspace{5mm} 
+ a_-^2\left(a_+-a_-\right)\ee^{-3H_0t} + \frac{a_-\left(a_+-a_-\right)^2}{2}\ee^{-6H_0t} + \frac{\left(a_+-a_-\right)^3}{9}\ee^{-9H_0t} {\Bigg]}  \, ,
\eea 
where only the $(k\eta)^2$-suppressed term that becomes dominant at late times has been kept, \ie there are other $(k\eta)^2$ terms that have been dropped for consistency since they always provide sub-dominant contributions. One can check that at early time, \ie when $t\to 0$, the ultra-slow-roll expression~(\ref{eq:alpha:USR:nu=3/2}) is recovered if $a_-/a_+\ll 1$, while at late time, \ie when $t \to \infty$, the slow-roll expression~(\ref{eq:alphaLO:SR}) is recovered. This gives rise to the gauge correction
\bea
\label{eq:Staro:gaugeCorr}
\widetilde{\delta\phi}_k/ Q_k & \kern-0.1em = \kern-0.1em 1 \kern-0.2em + \kern-0.2em \frac{1}{6}\left(\frac{H_0}{H}\right)^3\left[\left(a_--a_+\right)\ee^{-3H_0 t}-a_-\right]\kern-0.2em
{\Bigg[}  
\frac{(k\eta)^2}{2}a_- 
+ \left(a_--a_+\right)\ee^{-3H_0t}\left(1+\frac{a_-^2}{3}\right) 
\\ &
+ a_-^2\left(a_+-a_-\right)\ee^{-3H_0t} + \frac{a_-\left(a_+-a_-\right)^2}{2}\ee^{-6H_0t} + \frac{\left(a_+-a_-\right)^3}{9}\ee^{-9H_0t} {\Bigg]}\, ,
\eea
which is displayed in the right panel of \Fig{fig:Staro} for a mode such that $k/aH=10^{-2}$ at the transition time $t=0$. Right after the transition point, one can check that the ultra-slow-roll result~\eqref{eq:deltaphi:USRtransform} is recovered (the slight discrepancy is due to the finite value of $a_-/a_+$, \ie the finite initial value of $\delta$, we work with in \Fig{fig:Staro}), and at late time, the slow-roll result~\eqref{eq:GaugeCorr:SR} is obtained. In between, the gauge correction to the noise correlators remains tiny and can therefore be safely neglected.
\section{Conclusions}
\label{sec:conclusions}
In this paper we have discussed the challenges associated with using the stochastic formalism for inflation beyond the usual slow-roll approximation. One of the main pillars that stochastic inflation rests on is the separate universe approach, which pictures the universe on super-Hubble scales as causally disconnected regions that evolve under local FLRW dynamics. By showing that the dynamics of super-Hubble fluctuations can be recovered by perturbing the background FLRW equations of motion, we have demonstrated that this approach is valid and does not require slow roll. 

Subtleties however arise regarding the gauge in which the stochastic equations are written. 
The time variable we use in the Langevin equation is the logarithmic expansion, \ie the number of \efolds~$N$, and this variable is left unperturbed, which means we are implicitly working in a gauge in which the expansion is uniform.
However, the field fluctuations, which determine the correlations of the noises, are usually quantised in the spatially-flat gauge, where they coincide with the gauge-invariant Sasaki--Mukhanov scalar field perturbations, $Q$. 
One therefore has to perform a gauge transformation from the spatially-flat to the uniform-$N$ gauge before evaluating the stochastic noise due to quantum field fluctuations. We have calculated this transformation and shown that it is proportional to the non-adiabatic pressure perturbation. Since this vanishes on large scales in the presence of a dynamical attractor, such as in slow roll, the gauge transformation becomes trivial in that case (\ie the two gauges coincide on super-Hubble scales). We have also examined the case of ultra-slow roll, where we have found that the gauge transformation is also trivial, although ultra-slow roll is not a dynamical attractor in field space. Finally, we have studied the Starobinsky model, where the dynamics interpolates between a phase of ultra-slow-roll and slow-roll inflation, and have found that the same conclusions apply.

We conclude that in all three cases, the gauge transformation that is required prior to evaluating the noise correlators in stochastic inflation, turns out to be trivial on super-Hubble scales, and stochastic inflation as usually formulated can be applied without further refinements. This does not preclude the existence of situations where these gauge effects might be significant, but in such cases, \Eqs{eq:transform:phi}, \eqref{eq:GaugeTransf:pi}, \eqref{eq:sourcefunction:general} and \eqref{eq:alphaintegral1:general} provide the key formulas to compute them. This result has important consequences for the production of primordial black holes, whose formation is likely to require regimes of inflation that both violate slow roll and undergo large stochastic diffusion.
\acknowledgments

The authors are grateful to Hassan Firouzjahi and Enrico Pajer for useful discussions. 
HA and DW acknowledge support from the UK Science and Technology Facilities Council grants ST/N000668/1 and ST/S000550/1. 
VV acknowledges funding from the European Union's Horizon 2020 research and innovation programme under the Marie Sk\l odowska-Curie grant agreement N${}^0$ 750491.

\appendix 
\section{Sasaki--Mukhanov equation} \label{appendix:MSequation}
In this appendix, we  derive the general expression \eqref{eq:z''overz:general} for $z''/z$ in the Sasaki--Mukhanov equation and discuss both the slow-roll and the ultra-slow-roll limits.

\subsection*{Deriving the general expression}

We start with the Sasaki--Mukhanov variable
\bea \label{eq:def:v}
v_k = z\zeta_k \, ,
\eea 
where 
\bea \label{eq:def:z}
z = a \sqrt{2 \epsilon_{1}}\Mp \, ,
\eea 
and $v_k$ obeys the Sasaki--Mukhanov equation
\bea 
v_k'' + \left( k^2 - \frac{z''}{z} \right) v_k = 0 \, .
\eea 
Note that in the spatially flat gauge, we have $v_k = a \delta \phi_k =aQ_k $.
Let us also reiterate the notation we use: $\dot{ } = \frac{\dd}{\dd t}$ ($t$ is proper time) and $' = \frac{\dd}{\dd \eta}$ ($\eta$ is conformal time), so $\frac{\dd}{\dd \eta} = a \frac{\dd}{\dd t}$.
Combining
\bea 
\epsilon_{1} = -\frac{\dot{H}}{H^2} = 1 - \frac{\mathcal{H}'}{\mathcal{H}^2}
\eea 
with the Friedmann equation~(\ref{eq:friedmann}), one obtains $\epsilon_{1} =  \dot{\phi}^2/(2\Mp^2H^2)$, and $z$ can be rewritten as 
\bea 
z  = \frac{a \dot{\phi}}{H} \, .
\eea 
This allows us to calculate 
\bea \label{eq:z'/z}
\frac{z'}{z} = \frac{a}{z}\dot{z} = a\left( \frac{\dot{a}}{a} + \frac{\ddot{\phi}}{\dot{\phi}} - \frac{\dot{H}}{H} \right) \, .
\eea 
In terms of the slow-roll parameter $\epsilon_{1}$ and the relative acceleration parameter $f$ defined in \Eq{eq:def:f}, this reads 
\bea \label{eq:z'/z:exact}
\frac{z'}{z}= aH\left(1 - 3f + \epsilon_{1} \right) \, .
\eea 
In order to calculate $z''/z$, notice that
\bea \label{eq:intermediate}
\left(\frac{z'}{z}\right)' = \frac{z''}{z} - \frac{z'^2}{z^2} \, ,
\eea 
where the left-hand side can be derived from \Eq{eq:z'/z:exact},
\bea \label{eq:intermediate2}
\left(\frac{z'}{z}\right)' = a\frac{\dd}{\dd t}\left( \frac{z'}{z} \right) = a^2H\left(1 - 3f + \epsilon_{1} \right) \left( \frac{\dot{a}}{a} + \frac{\dot{H}}{H} + \frac{ \dot{\epsilon_{1}}-3\dot{f}}{1 - 3f + \epsilon_{1}} \right) \, .
\eea 
Derivating $\epsilon_{1}  = \dot{\phi}^2/(2\Mp^2H^2)$ with respect to time, and making use of the Klein--Gordon equation~(\ref{eq:kleingordon}), we have
\bea \label{eq:epsilondot}
\dot{\epsilon_{1}} = 2H\epsilon_{1} \left( \epsilon_{1} - 3f \right) \, ,
\eea 
and from \Eq{eq:def:f} we can calculate
\bea 
\frac{\dot{f}}{H} &= \mu + \left(f - 1\right)\left( \epsilon_{1} + 3f \right) \, .
\eea 
where the dimensionless mass parameter $\mu$ is defined in \Eq{mu:def}. Thus we can evaluate \Eq{eq:intermediate2} as
\bea 
\left(\frac{z'}{z}\right)' &= a^2H^2 \left( 1+6f+3\epsilon_{1} - 3\mu + \epsilon_{1}^2 - 9f^2 - 6f\epsilon_{1} \right) \, .
\eea 
Combining this with \Eqs{eq:z'/z:exact} and \eqref{eq:intermediate}, we find
\bea \label{eq:z''}
\frac{z''}{z} &= a^2H^2 \left( 2 + 5\epsilon_{1} - 3\mu - 12f\epsilon_{1} + 2\epsilon_{1}^2 \right)\, ,
\eea 
which is an exact expression that makes no approximations. 
\subsection*{Slow-roll limit}
\label{sec:z''/z:SR}
In the slow-roll regime,
\begin{align}
\label{eq:eps1:V}
\epsilon_{1}&\simeq \epsilon_{1}^V \equiv \frac{\Mp^2}{2} \left( \frac{V_{,\phi}}{V} \right)^2 \, , \\
 \label{eq:eps2:V}
\epsilon_{2}&\simeq \epsilon_{2}^V \equiv 2\Mp^2\left[ \left( \frac{V_{,\phi}}{V} \right)^2 - \frac{V_{,\phi\phi}}{V} \right] \, , \\
H^2 &\simeq \frac{V}{3\Mp^2} \, ,
\end{align}
and hence
\begin{align}
\mu &\simeq \Mp^2 \frac{V_{,\phi\phi}}{V} \\
\epsilon_{2} &\simeq 4\epsilon_{1} - 2 \mu \, .
\end{align}
At leading order in the slow-roll parameters, we therefore see that \eqref{eq:z''} reduces to 
\bea 
\frac{z''}{z} &\simeq a^2H^2 \left[ 2 - \epsilon_{1} + \frac{3}{2}\epsilon_{2}  +\order{\epsilon^2}\right] \, , 
\eea 
where we note that, since $f = \frac{2\epsilon_1-\epsilon_2}{6} $ \cite{Pattison:2018bct}, terms of order $\mathcal{O}(f\epsilon_{1})$ are neglected at first order.
In order to write \eqref{eq:z''} as an explicit function of conformal time, note that 
\bea \label{eq:eta:integral}
\eta &= \int \frac{\dd t}{a} = \int \frac{\dd a}{a^2H(a)} = -\frac{1}{aH} + \int \frac{\epsilon_{1} \dd a}{a^2H} \, ,
\eea 
where we have integrated by parts to get the last equality.
From \Eq{eq:epsilondot}, we have
\bea 
\frac{\dot{\epsilon_{1}}}{\epsilon_{1}} &= 2H \left( \epsilon_{1} - 3f \right) \, ,
\eea 
and thus we can again integrate by parts to find 
\bea 
\int \frac{\epsilon_{1} \dd a}{a^2H} &= -\frac{\epsilon_{1}}{aH} + \int \frac{\dd a}{a^2H}\frac{\dot{\epsilon_{1}}}{\epsilon_{1}}\frac{\epsilon_{1}}{H} + \mathcal{O}\left( \epsilon_{1}^2 \right)
 = -\frac{\epsilon_{1}}{aH} + \mathcal{O}\left(\epsilon_{1}^2, f\epsilon_{1} \right) \, .
\eea 
Therefore, from \Eq{eq:eta:integral},
\bea \label{eq:tau:slowroll}
\eta \simeq -\frac{1}{aH}\left( 1 + \epsilon_{1} \right)
\eea 
at first order in slow roll, and \Eq{eq:z''} becomes
\bea \label{eq:z'':slowrolllimit}
\frac{z''}{z} &\simeq \frac{2}{\eta^2} \left( 1 + \frac{3}{2}\epsilon_{1} + \frac{3}{4}\epsilon_{2} \right)\, ,
\eea 
in agreement with the well-known, leading-order slow-roll result.

\subsection*{Near ultra-slow-roll limit}
In the ultra-slow-roll regime, the field acceleration parameter is close to one and it is convenient to parameterise
\bea 
f = 1 - \delta \, ,
\eea 
where $\vert \delta \vert \ll 1$. In terms of $\delta$, \Eq{eq:z''} becomes
\bea 
\frac{z''}{z} &= a^2H^2 \left( 2 - 7\epsilon_{1} + 2 \epsilon_{1}^2 + 12\delta\epsilon_{1} - 3\mu \right) \, .
\eea 
In order to derive an explicit expression in terms of the conformal $\eta$ using \Eq{eq:eta:integral}, we note from \Eq{eq:epsilondot} that we have
\bea 
\frac{\dot{\epsilon_{1}}}{\epsilon_{1}} &= 2H \left( \epsilon_{1} - 3f \right)
 = -6H \left( 1 - \delta - \frac{\epsilon_{1}}{3} \right) \, .
\eea 
This allows us to again integrate by parts in \Eq{eq:eta:integral} and find
\bea \label{eq:integratebyparts}
\int \frac{\epsilon_{1} \dd a}{a^2H} &= -\frac{1}{7}\frac{\epsilon_{1}}{aH} + \mathcal{O}\left( \epsilon_{1}\delta, \epsilon_{1}^2 \right) \, ,
\eea 
and hence
\bea \label{eq:tau:USR}
\eta = -\frac{1}{aH}\left( 1 + \frac{1}{7}\epsilon_{1} \right)  + \mathcal{O}\left( \epsilon_{1}\delta, \epsilon_{1}^2 \right) \, .
\eea 
Thus, in ultra-slow roll, \Eq{eq:z''} becomes
\bea 
\label{eq:z''/z:usr}
\frac{z''}{z} & = \frac{1}{\eta^2} \left[ 2 - 3 \mu - \frac{3}{7}\epsilon_{1}\left( 15 + 2\mu \right) \right] + \mathcal{O}\left( \epsilon_{1}\delta, \epsilon_{1}^2 \right) \, .
\eea 
If the effective mass parameter $\mu$ is small, the leading-order behaviour is the same as in conventional slow roll, but it differs if $\mu$ is of order one or larger.
\subsection*{Solution in the slow-roll limit}
\label{appendix:solving:MSequation}
Let us start with \Eq{eq:z'':slowrolllimit}, and rewrite this as 
\bea \label{eq:z:nu}
\frac{z''}{z} \equiv \frac{\nu^2 - \frac{1}{4}}{\eta^2} \, ,
\eea 
where $\nu^2 =9/4 + 3\epsilon_{1*} + 3\epsilon_{2*}/2$ can be taken as constant at leading order in slow roll. At that order, the Sasaki--Mukhanov equation has the generic solution
\bea \label{eq:v:generalsolution:slowroll}
v_{k}\left(\eta\right) &= \sqrt{-\eta} \left[ A J_{\nu}\left(-k\eta \right) + BY_{\nu} \left( -k\eta \right) \right] \\
&= \sqrt{-\eta} \left[ \alpha H^{(1)}_{\nu}\left(-k\eta \right) + \beta H^{(2)}_{\nu} \left( -k\eta \right) \right] \, ,
\eea 
where conformal time $\eta$ runs from $-\infty$ to $0$ during inflation. 
In \Eq{eq:v:generalsolution:slowroll}, $J_{\nu}$ is the Bessel function of the first kind,  $Y_{\nu}$ is the Bessel function of the second kind, $A$, $B$, $\alpha$ and $\beta$ are constants, and the second line follows from 
\begin{align}
H_{\nu}^{(1)} &= J_{\nu} + iY_{\nu}\, , \\
H_{\nu}^{(2)} &= J_{\nu} - iY_{\nu} \, ,
\end{align}
where $H_{\nu}^{(1)}$ is the Hankel function of the first kind, and $H_{\nu}^{(2)}$ is the Hankel function of the second kind.

In order to fix the constants $A$ and $B$, or $\alpha$ and $\beta$, we set our initial conditions in the Bunch-Davies vacuum
\bea 
\lim_{\eta \to - \infty} v_{k}(\eta) = \frac{\ee^{-ik\eta}}{\sqrt{2k}}  \, .
\eea 
We implement this initial condition by making use of the following asymptotic behaviour for the Hankel functions 
\bea 
\lim_{k\eta \to -\infty} H_{\nu}^{(1)}\left(-k\eta \right) &= \sqrt{\frac{2}{\pi}}\frac{1}{\sqrt{-k\eta}} \ee^{ik\eta} \ee^{i \frac{\pi}{2} \left(\nu + \frac{1}{2}\right)} \\
\lim_{k\eta \to -\infty} H_{\nu}^{(2)}\left(-k\eta \right) &= \sqrt{\frac{2}{\pi}}\frac{1}{\sqrt{-k\eta}} \ee^{- ik\eta} \ee^{-i \frac{\pi}{2} \left(\nu + \frac{1}{2}\right)} \, .
\eea 
Thus 
\bea 
\lim_{k\eta \to -\infty} v_{k}\left(\eta\right) &= \sqrt{\frac{2}{\pi k}} \left[ \alpha \ee^{i \frac{\pi}{2} \left(\nu + \frac{1}{2}\right)} \ee^{ik\eta} + \beta \ee^{-i \frac{\pi}{2} \left(\nu + \frac{1}{2}\right)}\ee^{-ik\eta} \right] = \frac{\ee^{-ik\eta}}{\sqrt{2k}}  \, .
\eea 
By comparing these two expressions, we conclude that $\alpha = 0$ and $\beta = \frac{\sqrt{\pi}}{2}$ (where the irrelevant phase factor $\ee^{-i \frac{\pi}{2} \left(\nu + \frac{1}{2}\right)}$ is dropped). 
Thus the Bunch-Davies modes at first order in slow roll are
\bea 
v_{k} \left( \eta \right) = \frac{\sqrt{-\pi\eta}}{2} H_{\nu}^{(2)}\left(-k\eta \right) \, .
\eea 
\section{First slow-roll correction in ultra-slow roll}
\label{app:usr:epscorrections}
In this section, we solve \Eq{eq:MSequn} perturbatively in $\epsilon_{1*}$, in order to check that the leading-order solutions given in \Sec{sec:USR} are indeed consistent. We still consider the case of an exactly flat potential, so that $\delta=\mu=0$, and \Eq{eq:z''/z:usr} reads
\bea 
\frac{z''}{z} = \frac{1}{\eta^2}\left( 2 - \frac{45}{7}\epsilon_{1}\right) = \frac{1}{\eta^2}\left[  2 - \frac{45}{7}\epsilon_{1*}\left(\frac{\eta}{\eta_{*}}\right)^6\right] \, .
\eea
At first order in $\epsilon_{1*}$, the comoving Hubble parameter is given by \Eq{eq:tau:USR}, namely
\bea \label{eq:mathcalH:USR}
\mathcal{H} = -\frac{1}{\eta}\left[ 1 + \frac{1}{7}\epsilon_{1} + \mathcal{O}(\epsilon_{1}^2) \right] \, .
\eea 
Unlike the slow-roll case, this cannot be integrated to find $a(\eta)$, but we can instead perform an expansion in powers of $\epsilon_{1*}$ to find
\bea \label{eq:a:USR}
a(\eta) = -\frac{1}{H_* \eta}\left[ 1 - \frac{1}{42}\left(\epsilon_{1}-\epsilon_{1*}\right) + \mathcal{O}(\epsilon_{1}^2) \right] \, .
\eea

In order to solve \Eq{eq:MSequn} perturbatively, we first note that the leading order solution is simply the $\nu=\frac{3}{2}$ solution already given in \Eq{eq:v:nu=3/2}.
To find the first correction to this, we introduce 
\bea \label{eq:v:firstordereps}
v_{k} = \frac{\ee^{-ik\eta}}{\sqrt{2k}}\left(1 - \frac{i}{k\eta}\right)\left[ 1+\epsilon_{1*}f_k(\eta)\right] \, .
\eea
If we substitute \Eq{eq:v:firstordereps} back into \Eq{eq:MSequn} and solve the resultant differential equation for $f(\eta)$, we find that 
\bea \label{eq:f':usr}
f'_k(\eta) &=  -\frac{45}{28}\frac{1}{k^3}\frac{\eta^2}{\eta_{*}^6}\frac{\ee^{2ik\eta}}{(k\eta-i)^2}\left[ \ee^{-2ik\eta}\left(7i - 14k\eta - 14ik^2\eta^2+8k^3\eta^3 + 2ik^4\eta^4\right) \right. \\
& \hspace{2cm} \left. - \ee^{-2ik\eta_{\mathrm{start}}}\left(7i - 14k\eta_{\mathrm{start}} - 14ik^2\eta_{\mathrm{start}}^2+8k^3\eta_{\mathrm{start}}^3 + 2ik^4\eta_{\mathrm{start}}^4\right) \right] \, ,
\eea 
where the integration constant $\eta_{\mathrm{start}}$ must be chosen such that
\bea 
\epsilon_{1\mathrm{start}} = \epsilon_{1*}\left(\frac{\eta_{\mathrm{start}}}{\eta_{*}}\right)^6 < 1 \, .
\eea 
Combining \Eqs{eq:a:USR} and~(\ref{eq:v:firstordereps}) at leading order in $\epsilon_{1*}$, recalling that $v_k = a Q_k$, we can then calculate
\bea 
\frac{Q'_k}{Q_k} &=  - \frac{ik\eta}{\eta - \frac{i}{k}} + \epsilon_{1*}\left[ \frac{1}{7\eta}\left(\frac{\eta}{\eta_*}\right)^6 - f'_k\right] \, ,
\eea
where $f'$ is given by \Eq{eq:f':usr}.
Note that at leading order in $\epsilon_{1*}$, this reduces to \Eq{eq:Q'overQ:USR:nu=3over2}, as expected. We also see that the source function \eqref{eq:sourcefunction:general} becomes
\bea \label{eq:S:USReps1correction}
S_k &= - \sqrt{\frac{\epsilon_1}{2}}\frac{Q_k}{\Mp}\left\lbrace \frac{3}{\eta} - \frac{ik\eta}{\eta-\frac{i}{k}} - \epsilon_{1*}\left[ \frac{3}{7\eta}\left(\frac{\eta}{\eta_{*}}\right)^6  - f'_k\right] + \mathcal{O}(\epsilon_{1*}^2) \right\rbrace \, .
\eea
This implies that
\bea 
Q_k & = -\frac{H_*}{\sqrt{2k}}\ee^{-ik\eta}\left[ \eta - \frac{i}{k} + \mathcal{O}(\epsilon_{1*})\right] \\
S_k &= -\sqrt{\frac{\epsilon_{1*}}{2}}\frac{Q}{\Mp}\left(\frac{\eta}{\eta_*}\right)^3\left[ \frac{3}{\eta} - \frac{ik\eta}{\eta-\frac{i}{k}} + \mathcal{O}(\epsilon_{1*}) \right] \, .
\eea 
Making use of \Eq{eq:alphaintegral1:general}, we thus find 
\bea 
\alpha_k &
\simeq 
 -\frac{iH_{*}\sqrt{\epsilon_{1*}}}{6\Mp}{k^{-\frac{5}{2}}}(k\eta)^4
 \left[ 1 + \mathcal{O}(\epsilon_{1*})\right]   \, ,
\eea
of which \Eq{eq:alpha:USR:nu=3/2} indeed captures the leading order. The situation is therefore different than in slow roll where the leading order result vanishes and the dominant contribution comes from the decaying mode. This is because, as stressed above, the presence of a dynamical attractor in slow roll makes the non-adiabatic pressure perturbation vanish, which is not the case in ultra-slow roll.
\bibliographystyle{JHEP}
\bibliography{BeyondSlowRollStocha.bib}

\providecommand{\href}[2]{#2}\begingroup\raggedright\begin{thebibliography}{10}

\bibitem{Starobinsky:1980te}
A.~A. Starobinsky, \emph{{A New Type of Isotropic Cosmological Models Without
  Singularity}},
  \href{http://dx.doi.org/10.1016/0370-2693(80)90670-X}{\emph{Phys. Lett.} {\bf
  B91} (1980) 99--102}.

\bibitem{Sato:1980yn}
K.~Sato, \emph{{First Order Phase Transition of a Vacuum and Expansion of the
  Universe}}, {\emph{Mon.Not.Roy.Astron.Soc.} {\bf 195} (1981) 467--479}.

\bibitem{Guth:1980zm}
A.~H. Guth, \emph{{The Inflationary Universe: A Possible Solution to the
  Horizon and Flatness Problems}},
  \href{http://dx.doi.org/10.1103/PhysRevD.23.347}{\emph{Phys.Rev.} {\bf D23}
  (1981) 347--356}.

\bibitem{Linde:1981mu}
A.~D. Linde, \emph{{A New Inflationary Universe Scenario: A Possible Solution
  of the Horizon, Flatness, Homogeneity, Isotropy and Primordial Monopole
  Problems}},
  \href{http://dx.doi.org/10.1016/0370-2693(82)91219-9}{\emph{Phys.Lett.} {\bf
  B108} (1982) 389--393}.

\bibitem{Albrecht:1982wi}
A.~Albrecht and P.~J. Steinhardt, \emph{{Cosmology for Grand Unified Theories
  with Radiatively Induced Symmetry Breaking}},
  \href{http://dx.doi.org/10.1103/PhysRevLett.48.1220}{\emph{Phys.Rev.Lett.}
  {\bf 48} (1982) 1220--1223}.

\bibitem{Linde:1983gd}
A.~D. Linde, \emph{{Chaotic Inflation}},
  \href{http://dx.doi.org/10.1016/0370-2693(83)90837-7}{\emph{Phys.Lett.} {\bf
  B129} (1983) 177--181}.

\bibitem{Mukhanov:1981xt}
V.~F. Mukhanov and G.~Chibisov, \emph{{Quantum Fluctuation and Nonsingular
  Universe.}}, {\emph{JETP Lett.} {\bf 33} (1981) 532--535}.

\bibitem{Mukhanov:1982nu}
V.~F. Mukhanov and G.~Chibisov, \emph{{The Vacuum energy and large scale
  structure of the universe}}, {\emph{Sov.Phys.JETP} {\bf 56} (1982) 258--265}.

\bibitem{Starobinsky:1982ee}
A.~A. Starobinsky, \emph{{Dynamics of Phase Transition in the New Inflationary
  Universe Scenario and Generation of Perturbations}},
  \href{http://dx.doi.org/10.1016/0370-2693(82)90541-X}{\emph{Phys.Lett.} {\bf
  B117} (1982) 175--178}.

\bibitem{Guth:1982ec}
A.~H. Guth and S.~Pi, \emph{{Fluctuations in the New Inflationary Universe}},
  \href{http://dx.doi.org/10.1103/PhysRevLett.49.1110}{\emph{Phys.Rev.Lett.}
  {\bf 49} (1982) 1110--1113}.

\bibitem{Hawking:1982cz}
S.~Hawking, \emph{{The Development of Irregularities in a Single Bubble
  Inflationary Universe}},
  \href{http://dx.doi.org/10.1016/0370-2693(82)90373-2}{\emph{Phys.Lett.} {\bf
  B115} (1982) 295}.

\bibitem{Bardeen:1983qw}
J.~M. Bardeen, P.~J. Steinhardt and M.~S. Turner, \emph{{Spontaneous Creation
  of Almost Scale - Free Density Perturbations in an Inflationary Universe}},
  \href{http://dx.doi.org/10.1103/PhysRevD.28.679}{\emph{Phys.Rev.} {\bf D28}
  (1983) 679}.

\bibitem{Starobinsky:1986fx}
A.~A. Starobinsky, \emph{{Stochastic de Sitter (inflationary) stage in the
  early Universe}},
  \href{http://dx.doi.org/10.1007/3-540-16452-9_6}{\emph{Lect. Notes Phys.}
  {\bf 246} (1986) 107--126}.

\bibitem{Starobinsky:1994bd}
A.~A. Starobinsky and J.~Yokoyama, \emph{{Equilibrium state of a
  selfinteracting scalar field in the De Sitter background}},
  \href{http://dx.doi.org/10.1103/PhysRevD.50.6357}{\emph{Phys. Rev.} {\bf D50}
  (1994) 6357--6368}, [\href{http://arxiv.org/abs/astro-ph/9407016}{{\tt
  astro-ph/9407016}}].

\bibitem{Tsamis:2005hd}
N.~C. Tsamis and R.~P. Woodard, \emph{{Stochastic quantum gravitational
  inflation}},
  \href{http://dx.doi.org/10.1016/j.nuclphysb.2005.06.031}{\emph{Nucl. Phys.}
  {\bf B724} (2005) 295--328}, [\href{http://arxiv.org/abs/gr-qc/0505115}{{\tt
  gr-qc/0505115}}].

\bibitem{Finelli:2008zg}
F.~Finelli, G.~Marozzi, A.~A. Starobinsky, G.~P. Vacca and G.~Venturi,
  \emph{{Generation of fluctuations during inflation: Comparison of stochastic
  and field-theoretic approaches}},
  \href{http://dx.doi.org/10.1103/PhysRevD.79.044007}{\emph{Phys. Rev.} {\bf
  D79} (2009) 044007}, [\href{http://arxiv.org/abs/0808.1786}{{\tt
  0808.1786}}].

\bibitem{Garbrecht:2013coa}
B.~Garbrecht, G.~Rigopoulos and Y.~Zhu, \emph{{Infrared correlations in de
  Sitter space: Field theoretic versus stochastic approach}},
  \href{http://dx.doi.org/10.1103/PhysRevD.89.063506}{\emph{Phys. Rev.} {\bf
  D89} (2014) 063506}, [\href{http://arxiv.org/abs/1310.0367}{{\tt
  1310.0367}}].

\bibitem{Vennin:2015hra}
V.~Vennin and A.~A. Starobinsky, \emph{{Correlation Functions in Stochastic
  Inflation}},
  \href{http://dx.doi.org/10.1140/epjc/s10052-015-3643-y}{\emph{Eur. Phys. J.}
  {\bf C75} (2015) 413}, [\href{http://arxiv.org/abs/1506.04732}{{\tt
  1506.04732}}].

\bibitem{Onemli:2015pma}
V.~K. Onemli, \emph{{Vacuum Fluctuations of a Scalar Field during Inflation:
  Quantum versus Stochastic Analysis}},
  \href{http://dx.doi.org/10.1103/PhysRevD.91.103537}{\emph{Phys. Rev.} {\bf
  D91} (2015) 103537}, [\href{http://arxiv.org/abs/1501.05852}{{\tt
  1501.05852}}].

\bibitem{Burgess:2015ajz}
C.~P. Burgess, R.~Holman and G.~Tasinato, \emph{{Open EFTs, IR Effects and
  Late-Time Resummations: Systematic Corrections in Stochastic Inflation}},
  \href{http://arxiv.org/abs/1512.00169}{{\tt 1512.00169}}.

\bibitem{Vennin:2016wnk}
V.~Vennin, H.~Assadullahi, H.~Firouzjahi, M.~Noorbala and D.~Wands,
  \emph{{Critical Number of Fields in Stochastic Inflation}},
  \href{http://dx.doi.org/10.1103/PhysRevLett.118.031301}{\emph{Phys. Rev.
  Lett.} {\bf 118} (2017) 031301}, [\href{http://arxiv.org/abs/1604.06017}{{\tt
  1604.06017}}].

\bibitem{Hardwick:2017fjo}
R.~J. Hardwick, V.~Vennin, C.~T. Byrnes, J.~Torrado and D.~Wands, \emph{{The
  stochastic spectator}},
  \href{http://dx.doi.org/10.1088/1475-7516/2017/10/018}{\emph{JCAP} {\bf 1710}
  (2017) 018}, [\href{http://arxiv.org/abs/1701.06473}{{\tt 1701.06473}}].

\bibitem{Tokuda:2017fdh}
J.~Tokuda and T.~Tanaka, \emph{{Statistical nature of infrared dynamics on de
  Sitter background}},
  \href{http://dx.doi.org/10.1088/1475-7516/2018/02/014}{\emph{JCAP} {\bf 1802}
  (2018) 014}, [\href{http://arxiv.org/abs/1708.01734}{{\tt 1708.01734}}].

\bibitem{Garcia-Bellido:2017mdw}
J.~Garcia-Bellido and E.~Ruiz~Morales, \emph{{Primordial black holes from
  single field models of inflation}},
  \href{http://arxiv.org/abs/1702.03901}{{\tt 1702.03901}}.

\bibitem{Germani:2017bcs}
C.~Germani and T.~Prokopec, \emph{{On primordial black holes from an inflection
  point}},  \href{http://arxiv.org/abs/1706.04226}{{\tt 1706.04226}}.

\bibitem{Firouzjahi:2018vet}
H.~Firouzjahi, A.~Nassiri-Rad and M.~Noorbala, \emph{{Stochastic Ultra Slow
  Roll Inflation}},  \href{http://arxiv.org/abs/1811.02175}{{\tt 1811.02175}}.

\bibitem{Biagetti:2018pjj}
M.~Biagetti, G.~Franciolini, A.~Kehagias and A.~Riotto, \emph{{Primordial Black
  Holes from Inflation and Quantum Diffusion}},
  \href{http://arxiv.org/abs/1804.07124}{{\tt 1804.07124}}.

\bibitem{Ezquiaga:2018gbw}
J.~M. Ezquiaga and J.~Garcia-Bellido, \emph{{Quantum diffusion beyond
  slow-roll: implications for primordial black-hole production}},
  \href{http://arxiv.org/abs/1805.06731}{{\tt 1805.06731}}.

\bibitem{Pattison:2017mbe}
C.~Pattison, V.~Vennin, H.~Assadullahi and D.~Wands, \emph{{Quantum diffusion
  during inflation and primordial black holes}},
  \href{http://dx.doi.org/10.1088/1475-7516/2017/10/046}{\emph{JCAP} {\bf 1710}
  (2017) 046}, [\href{http://arxiv.org/abs/1707.00537}{{\tt 1707.00537}}].

\bibitem{Inoue:2001zt}
S.~Inoue and J.~Yokoyama, \emph{{Curvature perturbation at the local extremum
  of the inflaton's potential}},
  \href{http://dx.doi.org/10.1016/S0370-2693(01)01369-7}{\emph{Phys. Lett.}
  {\bf B524} (2002) 15--20}, [\href{http://arxiv.org/abs/hep-ph/0104083}{{\tt
  hep-ph/0104083}}].

\bibitem{Kinney:2005vj}
W.~H. Kinney, \emph{{Horizon crossing and inflation with large eta}},
  \href{http://dx.doi.org/10.1103/PhysRevD.72.023515}{\emph{Phys. Rev.} {\bf
  D72} (2005) 023515}, [\href{http://arxiv.org/abs/gr-qc/0503017}{{\tt
  gr-qc/0503017}}].

\bibitem{Pattison:2018bct}
C.~Pattison, V.~Vennin, H.~Assadullahi and D.~Wands, \emph{{The attractive
  behaviour of ultra-slow-roll inflation}},
  \href{http://dx.doi.org/10.1088/1475-7516/2018/08/048}{\emph{JCAP} {\bf 1808}
  (2018) 048}, [\href{http://arxiv.org/abs/1806.09553}{{\tt 1806.09553}}].

\bibitem{Cruces:2018cvq}
D.~Cruces, C.~Germani and T.~Prokopec, \emph{{Failure of the stochastic
  approach to inflation in constant-roll and ultra-slow-roll}},
  \href{http://arxiv.org/abs/1807.09057}{{\tt 1807.09057}}.

\bibitem{Grain:2017dqa}
J.~Grain and V.~Vennin, \emph{{Stochastic inflation in phase space: Is slow
  roll a stochastic attractor?}},
  \href{http://dx.doi.org/10.1088/1475-7516/2017/05/045}{\emph{JCAP} {\bf 1705}
  (2017) 045}, [\href{http://arxiv.org/abs/1703.00447}{{\tt 1703.00447}}].

\bibitem{Polarski:1995jg}
D.~Polarski and A.~A. Starobinsky, \emph{{Semiclassicality and decoherence of
  cosmological perturbations}},
  \href{http://dx.doi.org/10.1088/0264-9381/13/3/006}{\emph{Class. Quant.
  Grav.} {\bf 13} (1996) 377--392},
  [\href{http://arxiv.org/abs/gr-qc/9504030}{{\tt gr-qc/9504030}}].

\bibitem{Lesgourgues:1996jc}
J.~Lesgourgues, D.~Polarski and A.~A. Starobinsky, \emph{{Quantum to classical
  transition of cosmological perturbations for nonvacuum initial states}},
  \href{http://dx.doi.org/10.1016/S0550-3213(97)00224-1}{\emph{Nucl. Phys.}
  {\bf B497} (1997) 479--510}, [\href{http://arxiv.org/abs/gr-qc/9611019}{{\tt
  gr-qc/9611019}}].

\bibitem{Kiefer:2008ku}
C.~Kiefer and D.~Polarski, \emph{{Why do cosmological perturbations look
  classical to us?}}, \href{http://dx.doi.org/10.1166/asl.2009.1023}{\emph{Adv.
  Sci. Lett.} {\bf 2} (2009) 164--173},
  [\href{http://arxiv.org/abs/0810.0087}{{\tt 0810.0087}}].

\bibitem{Martin:2015qta}
J.~Martin and V.~Vennin, \emph{{Quantum Discord of Cosmic Inflation: Can we
  Show that CMB Anisotropies are of Quantum-Mechanical Origin?}},
  \href{http://dx.doi.org/10.1103/PhysRevD.93.023505}{\emph{Phys. Rev.} {\bf
  D93} (2016) 023505}, [\href{http://arxiv.org/abs/1510.04038}{{\tt
  1510.04038}}].

\bibitem{2005PhRvA..71b2103R}
M.~{Revzen}, P.~A. {Mello}, A.~{Mann} and L.~M. {Johansen}, \emph{{Bell's
  inequality violation with non-negative Wigner functions}},
  \href{http://dx.doi.org/10.1103/PhysRevA.71.022103}{\emph{Physical Review A}
  {\bf 71} (Feb, 2005) 022103},
  [\href{http://arxiv.org/abs/quant-ph/0405100}{{\tt quant-ph/0405100}}].

\bibitem{Martin:2016tbd}
J.~Martin and V.~Vennin, \emph{{Bell inequalities for continuous-variable
  systems in generic squeezed states}},
  \href{http://dx.doi.org/10.1103/PhysRevA.93.062117}{\emph{Phys. Rev.} {\bf
  A93} (2016) 062117}, [\href{http://arxiv.org/abs/1605.02944}{{\tt
  1605.02944}}].

\bibitem{Martin:2017zxs}
J.~Martin and V.~Vennin, \emph{{Obstructions to Bell CMB Experiments}},
  \href{http://dx.doi.org/10.1103/PhysRevD.96.063501}{\emph{Phys. Rev.} {\bf
  D96} (2017) 063501}, [\href{http://arxiv.org/abs/1706.05001}{{\tt
  1706.05001}}].

\bibitem{Salopek:1990jq}
D.~S. Salopek and J.~R. Bond, \emph{{Nonlinear evolution of long wavelength
  metric fluctuations in inflationary models}},
  \href{http://dx.doi.org/10.1103/PhysRevD.42.3936}{\emph{Phys. Rev.} {\bf D42}
  (1990) 3936--3962}.

\bibitem{Sasaki:1995aw}
M.~Sasaki and E.~D. Stewart, \emph{{A General analytic formula for the spectral
  index of the density perturbations produced during inflation}},
  \href{http://dx.doi.org/10.1143/PTP.95.71}{\emph{Prog.Theor.Phys.} {\bf 95}
  (1996) 71--78}, [\href{http://arxiv.org/abs/astro-ph/9507001}{{\tt
  astro-ph/9507001}}].

\bibitem{Wands:2000dp}
D.~Wands, K.~A. Malik, D.~H. Lyth and A.~R. Liddle, \emph{{A New approach to
  the evolution of cosmological perturbations on large scales}},
  \href{http://dx.doi.org/10.1103/PhysRevD.62.043527}{\emph{Phys.Rev.} {\bf
  D62} (2000) 043527}, [\href{http://arxiv.org/abs/astro-ph/0003278}{{\tt
  astro-ph/0003278}}].

\bibitem{Lyth:2003im}
D.~H. Lyth and D.~Wands, \emph{{Conserved cosmological perturbations}},
  \href{http://dx.doi.org/10.1103/PhysRevD.68.103515}{\emph{Phys.Rev.} {\bf
  D68} (2003) 103515}, [\href{http://arxiv.org/abs/astro-ph/0306498}{{\tt
  astro-ph/0306498}}].

\bibitem{Rigopoulos:2003ak}
G.~I. Rigopoulos and E.~P.~S. Shellard, \emph{{The separate universe approach
  and the evolution of nonlinear superhorizon cosmological perturbations}},
  \href{http://dx.doi.org/10.1103/PhysRevD.68.123518}{\emph{Phys. Rev.} {\bf
  D68} (2003) 123518}, [\href{http://arxiv.org/abs/astro-ph/0306620}{{\tt
  astro-ph/0306620}}].

\bibitem{Lyth:2005fi}
D.~H. Lyth and Y.~Rodriguez, \emph{{The Inflationary prediction for primordial
  non-Gaussianity}},
  \href{http://dx.doi.org/10.1103/PhysRevLett.95.121302}{\emph{Phys.Rev.Lett.}
  {\bf 95} (2005) 121302}, [\href{http://arxiv.org/abs/astro-ph/0504045}{{\tt
  astro-ph/0504045}}].

\bibitem{Lifshitz:1960}
E.~M. Lifshitz and I.~M. Khalatnikov, \emph{{About singularities of
  cosmological solutions of the gravitational equations. I}}, {\emph{ZhETF}
  {\bf 39} (1960) 149}.

\bibitem{Starobinsky:1982mr}
A.~A. Starobinsky, \emph{{Isotropization of arbitrary cosmological expansion
  given an effective cosmological constant}}, {\emph{JETP Lett.} {\bf 37}
  (1983) 66--69}.

\bibitem{Comer:1994np}
G.~Comer, N.~Deruelle, D.~Langlois and J.~Parry, \emph{{Growth or decay of
  cosmological inhomogeneities as a function of their equation of state}},
  \href{http://dx.doi.org/10.1103/PhysRevD.49.2759}{\emph{Phys.Rev.} {\bf D49}
  (1994) 2759--2768}.

\bibitem{Khalatnikov:2002kn}
I.~Khalatnikov, A.~Y. Kamenshchik and A.~A. Starobinsky, \emph{{Comment about
  quasiisotropic solution of Einstein equations near cosmological
  singularity}},
  \href{http://dx.doi.org/10.1088/0264-9381/19/14/322}{\emph{Class.Quant.Grav.}
  {\bf 19} (2002) 3845--3850}, [\href{http://arxiv.org/abs/gr-qc/0204045}{{\tt
  gr-qc/0204045}}].

\bibitem{Fujita:2013cna}
T.~Fujita, M.~Kawasaki, Y.~Tada and T.~Takesako, \emph{{A new algorithm for
  calculating the curvature perturbations in stochastic inflation}},
  \href{http://dx.doi.org/10.1088/1475-7516/2013/12/036}{\emph{JCAP} {\bf 1312}
  (2013) 036}, [\href{http://arxiv.org/abs/1308.4754}{{\tt 1308.4754}}].

\bibitem{Gordon:2000hv}
C.~Gordon, D.~Wands, B.~A. Bassett and R.~Maartens, \emph{{Adiabatic and
  entropy perturbations from inflation}},
  \href{http://dx.doi.org/10.1103/PhysRevD.63.023506}{\emph{Phys. Rev.} {\bf
  D63} (2001) 023506}, [\href{http://arxiv.org/abs/astro-ph/0009131}{{\tt
  astro-ph/0009131}}].

\bibitem{Malik:2008im}
K.~A. Malik and D.~Wands, \emph{{Cosmological perturbations}},
  \href{http://dx.doi.org/10.1016/j.physrep.2009.03.001}{\emph{Phys. Rept.}
  {\bf 475} (2009) 1--51}, [\href{http://arxiv.org/abs/0809.4944}{{\tt
  0809.4944}}].

\bibitem{Sasaki:1986hm}
M.~Sasaki, \emph{{Large Scale Quantum Fluctuations in the Inflationary
  Universe}}, \href{http://dx.doi.org/10.1143/PTP.76.1036}{\emph{Prog. Theor.
  Phys.} {\bf 76} (1986) 1036}.

\bibitem{Mukhanov:1988jd}
V.~F. Mukhanov, \emph{{Quantum Theory of Gauge Invariant Cosmological
  Perturbations}}, {\emph{Sov. Phys. JETP} {\bf 67} (1988) 1297--1302}.

\bibitem{Starobinsky:1992ts}
A.~A. Starobinsky, \emph{{Spectrum of adiabatic perturbations in the universe
  when there are singularities in the inflation potential}}, {\emph{JETP Lett.}
  {\bf 55} (1992) 489--494}.

\bibitem{Martin:2011sn}
J.~Martin and L.~Sriramkumar, \emph{{The scalar bi-spectrum in the Starobinsky
  model: The equilateral case}},
  \href{http://dx.doi.org/10.1088/1475-7516/2012/01/008}{\emph{JCAP} {\bf 1201}
  (2012) 008}, [\href{http://arxiv.org/abs/1109.5838}{{\tt 1109.5838}}].

\end{thebibliography}\endgroup
\end{document}